\documentclass[conference]{IEEEtran}

\usepackage{amsmath,bm}
\usepackage[ruled,vlined]{algorithm2e}

\usepackage{color}
\usepackage{enumitem}
\usepackage{soul,xcolor}
\usepackage[utf8]{inputenc}
\usepackage{amsmath,amssymb,graphicx}
\usepackage{derivative}
\usepackage{subfig}
\usepackage{tablefootnote}
\usepackage{commath}
\usepackage{lettrine}
\usepackage[inner=0.675in,outer=0.675in,top=0.7in,bottom=1.00in,pdftex]{geometry}
\usepackage[protrusion=true,expansion=true]{microtype}

\newcommand{\resolved}[1]{{\color{green} {\bf[RESOLVED]}}}
\DeclareMathOperator{\E}{\mathbb{E}}
\DeclareMathOperator{\R}{\mathbb{R}}
\DeclareMathOperator*{\argmin}{arg\,min}
\DeclareMathOperator*{\argmax}{arg\,max}

\allowdisplaybreaks

\begin{document}
\title{Resource-Efficient and Delay-Aware Federated Learning Design under Edge Heterogeneity}

\author{\IEEEauthorblockN{David Nickel\IEEEauthorrefmark{1}, Frank Po-Chen Lin\IEEEauthorrefmark{1}, Seyyedali Hosseinalipour\IEEEauthorrefmark{1}, Nicolo Michelusi\IEEEauthorrefmark{2}, and Christopher G. Brinton\IEEEauthorrefmark{1}}
\IEEEauthorblockA{\IEEEauthorrefmark{1}Electrical and Computer Engineering, Purdue University, West Lafayette, IN, USA}
\IEEEauthorblockA{\IEEEauthorrefmark{2}Electrical, Computer, and Energy Engineering, Arizona State University, Tempe, AZ, USA}
\IEEEauthorblockA{\IEEEauthorrefmark{1}\{dnickel, lin1183, hosseina, cgb\}@purdue.edu, \IEEEauthorrefmark{2}nicolo.michelusi@asu.edu}}


%


\maketitle

\begin{abstract}
Federated learning (FL) has emerged as a popular technique for distributing machine learning across wireless edge devices. We examine FL under two salient properties of contemporary networks: device-server communication delays and device computation heterogeneity. Our proposed {\tt StoFedDelAv} algorithm incorporates a local-global model combiner into the FL synchronization step. We theoretically characterize the convergence behavior of {\tt StoFedDelAv} and obtain the optimal \textit{combiner weights}, which consider the global model delay and expected local gradient error at each device. We then formulate a network-aware optimization problem which tunes the minibatch sizes of the devices to jointly minimize energy consumption and machine learning training loss, and solve the non-convex problem through a series of convex approximations. Our simulations reveal that {\tt StoFedDelAv} outperforms the current art in FL, evidenced by the obtained improvements in optimization objective.

\end{abstract}


%
\IEEEpeerreviewmaketitle

\section{Introduction} 
\vspace{-.5mm}
\noindent Recent advancements in smart devices (e.g. cell phones, smart cars) have resulted in a paradigm shift for machine learning (ML)~\cite{Chiang}, aiming to migrate intelligence management from cloud datacenters to the network edge~\cite{McMahan}. Federated learning (FL) has been promoted as one of the main frameworks for distributing ML over wireless networks~\cite{konevcny2016federated}, where model training is conducted without data exchange across devices. 

Conventional FL operates in two iterative steps~\cite{8851249}: (i) local training, where edge devices update their local models using their own datasets; and (ii) global aggregation, where a cloud server computes the global model based on local models received from the edge devices, and synchronizes them~\cite{wang2019adaptive}. Implementations of this process over the wireless edge are complicated by heterogeneity in communication and computation capabilities found across devices~\cite{hosseinalipour2020federated}. In this work, we augment FL to provide resilience to these factors.

\vspace{-1.5mm}
\subsection{Related Works}
\vspace{-.5mm}
Several works in FL have focused on techniques for improving device-to-server communication efficiency in the global aggregation step. Some have focused on reducing the number of uplink/downlink communication rounds by performing multiple iterations of local model updates between consecutive global aggregations~\cite{haddadpour2019convergence,tran2019federated}. Works~\cite{lin2021timescale,hosseinalipour2020multi} showed that device-server communication requirements in FL can be further reduced through direct device-to-device model synchronization.

Building upon this, there has been a recent trend towards control methodologies for optimizing device participation in FL. The authors of~\cite{9148815} proposed a joint optimization formulation considering learning, resource allocation, and device selection to minimize convergence time. In~\cite{yang2019energy}, the authors minimized the total energy consumption of the system under device heterogeneity constraints. In~\cite{yang2020federated}, the authors developed over-the-air FL for maximizing global model aggregation speed under proper device selection and beamforming design.

Such works have largely neglected the effect of \textit{communication delay} on the performance of model training in FL. In~\cite{frank2020delay}, we took a step towards addressing this by establishing a delay-aware FL framework. Specifically, we introduced a mechanism for devices to combine local and global models during the synchronization step to account for communication delay.
Nevertheless,~\cite{frank2020delay} considers a scenario in which the edge devices train their models in the local straining step using full-batch gradient descent (GD). This can introduce large inefficiencies with respect to the energy consumed versus model convergence obtained in FL, especially when training models over heterogeneous wireless devices. In practice, an edge device can potentially store more data than it can process in a timely manner. An energy-efficient solution to this is using minibatch stochastic gradient descent (SGD), which on the other hand has the downside of introducing estimation noise~\cite{lin2021timescale}. In this paper, we address these challenges by coupling the selection of device minibatch sizes with the weighting of  local and global model combiners based on heterogeneity conditions.

\vspace{-1.5mm}
\subsection{Outline and Summary of Contributions}
\vspace{-.5mm}
\vspace{-.3mm}
\begin{itemize}[leftmargin=5mm]
    \item We develop a delay-aware FL framework, {\tt StoFedDelAv}, which incorporates a local-global model combiner to jointly optimize model training performance and network resource consumption in the presence of device-server communication delays and device computation heterogeneity.
    \item We theoretically characterize the convergence behavior of {\tt StoFedDelAv} and  optimize the local-global model combiner weight in the presence of communication delay. We further formulate a network-aware learning optimization problem which aims to tune the SGD minibatch sizes across the devices according to resource constraints. We demonstrate that the problem is a non-convex signomial program, and solve it using a series of convex approximations.
    \item Our experiments show that {\tt StoFedDelAv} outperforms the current art in FL in terms of model convergence speed and network resource utilization when the minibatch size and local-global model combiner are carefully adjusted.
\end{itemize}

\section{System Model and Algorithm}\label{sec:sysMod}

\subsection{Network and Machine Learning Model}\label{ssec:network_model}
\begin{figure}[t]
\centering
\includegraphics[width=0.45\textwidth]{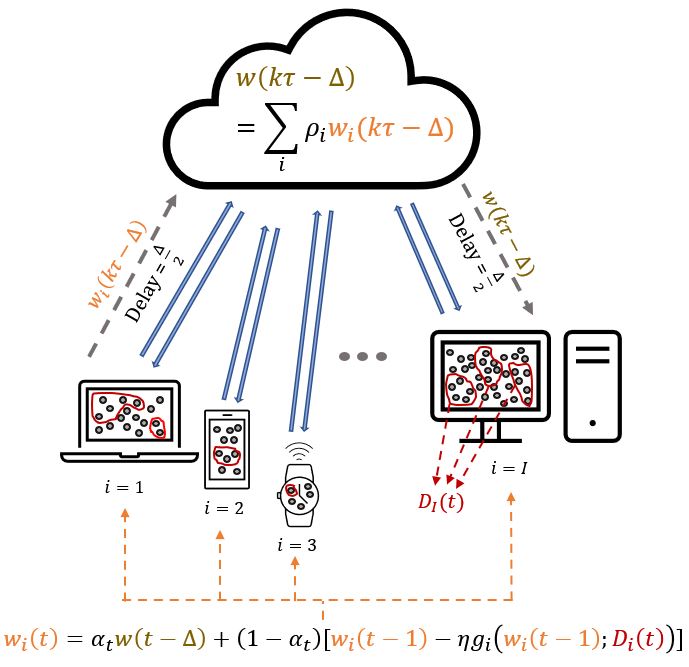}
\vspace{-2mm}
\caption{System architecture and illustration of our proposed methodology for delay-aware federated learning.}
\vspace{-0.25in}
\label{fig:sys_diagram}
\end{figure}

We consider a set $\mathcal{I} =\{1,\cdots, I\}$ of $I$ edge devices connected to a cloud server, which acts as a model aggregator (see Fig.~\ref{fig:sys_diagram}). Each edge device \textit{i} is associated with a dataset $\mathcal{D}_i$, where each datapoint $(\textbf{x},y)\in\mathcal{D}_i$ comprises an $m$-dimensional feature vector, $\textbf{x}\in\R^m$, and a label, $y\in\R$.

We let $f_i(\textbf{x},y;\textbf{w})$ be the loss of the machine learning model associated with datapoint $(\textbf{x},y)$ and model parameter vector $\textbf{w} \in \mathbb{R}^n$. The local loss function of device $i$ is given by
\vspace{-2mm}

{\small
\begin{equation}\label{LocLossDef}
	F_i(\textbf{w}) = \sum_{(\textbf{x},y)\in \mathcal{D}_i}{f_i(\textbf{x},y;\textbf{w})}/N_i.
\end{equation}}
\vspace{-4mm}

The global loss is subsequently defined as
\vspace{-2mm}

{\small
\begin{equation}\label{eq:globLossDef}
	F(\textbf{w}) = \sum_{i\in\mathcal{I}}{\rho_i F_i(\textbf{w})},
\end{equation}}
\vspace{-2mm}

\vspace{-1.5mm}
\noindent where $\rho_i = N_i / \sum_{j\in\mathcal{I}}{N_j}$ is the weight associated with device $i$. $N_{i}=|\mathcal{D}_i|$ is the size of the local dataset. The goal of the ML training is to find the optimal parameter given by
\vspace{-2mm}

{\small
\begin{equation}\label{eq:trng_goal}
	\textbf{w}^{\star}=\argmin_\textbf{w}{F(\textbf{w})}.
\end{equation}}
\vspace{-2mm}

\vspace{-3mm}
To aid in convergence analysis of model training across the network, the following assumptions are made:\\
\vspace{-3mm} 
\noindent\textbf{\\Assumption 1.} \textit{The loss functions are assumed to be L-Lipschitz and $\beta$-Smooth, i.e., $\|F_i(\textbf{w}_1)-F_i(\textbf{w}_2)\| \leq L\|\textbf{w}_1-\textbf{w}_2\|,\forall i$, and $\|\nabla F_i(\textbf{w}_1)-\nabla F_i(\textbf{w}_2)\|\leq \beta\|\textbf{w}_1-\textbf{w}_2\|,\forall i$.}

\noindent\textbf{Assumption 2.} \textit{The local and global gradients are assumed to have a bounded dissimilarity, i.e. $\|\nabla F_i(\textbf{w})-\nabla F(\textbf{w})\|\leq\delta_i,\forall\textbf{w},\forall i$,\noindent\textit{where $0 \leq \delta_{i} \leq 2L$. We let $\delta = \sum_{i} \rho_i \delta_{i}$.}}

\vspace{1mm}
Note that a higher value of $\delta$ implies a larger statistical diversity across the local datasets of the edge devices.
 
\vspace{-1mm}
\subsection{{\tt StoFedDelAv} Algorithm}\label{ssec:algo}
\vspace{-.5mm}
We propose the \texttt{StoFedDelAv} algorithm (see Alg. 1), considering the effect of the communication delay between the edge devices and the cloud server. 
We divide the full training cycle into discrete time-instances $t\in\{1,2,...,T\}$, where the training consists of $K=\frac{T}{\tau}$ rounds of aggregation. $\tau$ denotes the number of SGD steps taken by each device for each round of global aggregation indexed by $k\in\{0,1,...,K-1\}$, where each aggregation period spans the interval $\mathcal T_k = \{k\tau-\Delta+1,...,(k+1)\tau-\Delta\}.$ The communication delay, i.e., the duration between when edge devices send their models to the server and the  reception of the resulting global model is denoted by $\Delta$, where $\tau\geq\Delta\geq 0$. Without loss of generality, we assume the uplink and downlink communication delay to be symmetric, i.e., $\Delta / 2$, for both upstream and downstream communications.

Let $\mathbf w_i{(t)}$ denote the local model trained at each device $i$ and $\mathbf w(t) = \sum_i \rho_i \mathbf w_i(t)$ be the global model at each time instance $t$. The model training starts with the cloud server initializing all the local models such that $\mathbf w_i{(-\Delta)}={\mathbf{w}}{(-\Delta)}$, $\forall i$.

Between two consecutive global aggregations, each device sends its local model $\mathbf w_i(t)$ to the server at $t\in\{k\tau-\Delta,\forall k\geq 0\}$, after waiting for the communication delay between edge and server, i.e., $\Delta/2$, and the global model $\mathbf w(t)$ is computed at the server at $t\in\{k\tau-\Delta/2,\forall k\geq 0\}$. Finally, the devices receive the global model at $k\tau$ to perform local model synchronization.

\noindent\textbf{Distributed SGD:}
At time $t$, the edge devices sample their datasets randomly and without replacement, obtaining minibatch $\mathcal{D}_i(t)\subseteq\mathcal{D}_i$, where $|\mathcal{D}_i(t)|$ is the number of datapoints selected and is the same for each $t\in \mathcal T_k$. Let $n_i(k)\triangleq |\mathcal{D}_i(t)|,~\forall t\in\mathcal T_k$ be the minibatch size of device $i$ for the $k$-th aggregation period, each local device take an SGD step on their local model using unbiased gradient estimator as:

\vspace{-5mm}
{\small
\begin{align}\label{eq:SGD}
    g_i(\textbf{w}_i(t);\mathcal{D}_i(t))={\vert\mathcal{D}_i(t)\vert}^{-1}\sum_{(\mathbf x,y)\in\mathcal{D}_i(t)}
    \nabla f_i(\mathbf x,y;\mathbf w_i{(t)}),
\end{align}}

\vspace{-2.5mm}
\noindent where 
\vspace{-2mm}
\begin{equation}\label{eq:SGDwithNoise}
    g_i(\textbf{w}_i(t);\mathcal{D}_i(t))=\nabla F_i(\textbf{w}_i(t))+\nu_i(t)
\end{equation}

\vspace{-2.5mm}
\noindent with $\nu_i(t)$ being a zero-mean noise.

At each time $t \in \mathcal T_k \setminus\{k\tau\}$, each edge device updates the local model Using the gradient estimate as:
\vspace{-2.5mm}
\begin{align} \label{8}
    {{\mathbf{w}}}_i{(t)} = 
           \mathbf w_i{(t-1)}-\eta g_i(\textbf{w}_i(t);\mathcal{D}_i(t)),~t\in\mathcal T_k.
\end{align} 

\vspace{-3mm}
\noindent\textbf{Model Synchronization:}
At time $t=k\tau$, after receiving the delayed global model $\mathbf w(t-\Delta)$
from the cloud server, each edge device performs one additional local SGD update
followed by synchronization. During synchronization, each edge device performs local update by replacing its local model with a combination of the global and local model with the global/local {\it combiner weight} $\alpha(k)\in(0,1]$. The expression for the local model after synchronization is given by

\vspace{-4.2mm}
{\small
\begin{equation}\label{SGDstep}
    \begin{aligned}
        &\textbf{w}_i(t)=\alpha_t(k)\textbf{w}(t-\Delta)\\
        &+(1-\alpha_t(k))\left[\textbf{w}_i(t-1)-\eta g_i(\textbf{w}_{i}(t-1);\mathcal{D}_{i}(t))\right],
    \end{aligned}
\end{equation}}
\vspace{-4mm}

\noindent where $\alpha_t(k)$ is the weight assigned to the global model:

\vspace{-3mm}
{\small
\begin{equation}
    \alpha_t(k)=
    \begin{cases}
      \alpha(k), & t=k\tau,k\in\{0,1,...,K-1\} \\
      0, & \text{otherwise}
    \end{cases}.
\end{equation}}

\vspace{-3mm}
\noindent Let $\widehat{\bm{\alpha}}=\{\alpha(0),...,\alpha(K)\}$ be the set of {\it combiner weights} across the global aggregation instances. $\alpha=1$ corresponds to standard FL.

At the $K$-th global aggregation, the server chooses the best $\textbf{w}(t)$ it has found thus far. Since the server only has access to the global model at $t=k\tau-\Delta$, the model selected at $K$ is
\vspace{-3mm}

{\small
\begin{equation}\label{wKChosen}
    \textbf{w}^{K}=\min_{\textbf{w}\in\mathcal{W}}F(\textbf{w}),
\end{equation}} 

\vspace{-4mm}
\noindent with $\mathcal{W}\triangleq\{\textbf{w}(k\tau-\Delta),k=0,1,...,K-1\}$. 

{\footnotesize
\begin{algorithm} 
\SetAlgoLined
\caption{Stochastic Federated Delayed Averaging} \label{alg}
\KwIn{$\widehat{\bm{\alpha}},\tau,\mathcal{I},T$}
\KwOut{$\textbf{w}^{K}$}
 Initialize $\mathbf w_i(-\Delta),\  \forall i$\;
 \For{$k=0:K-1$}{
     \For{$t=k\tau-\Delta+1:(k+1)\tau-\Delta$}{
     \For{$i\in\mathcal{I}$}{
      \uIf{$t=(k+1)\tau-\Delta$}{
      Each device $i$ sends $\mathbf w_i$  to the server
       }
        \uElse{
            Device $i\in\mathcal{I}$ updates its model using \eqref{SGDstep}
        }
        }
        \uIf{$t=(k+1)\tau-\Delta/2$}{
        // Procedure at the cloud server \\
            Compute $\mathbf w((k+1)\tau-\Delta)$ and send it to the edge for synchronization and 
            update $\mathbf w^K$ with \eqref{wKChosen}
          }
     }
 }
\end{algorithm}}

\section{Convergence Analysis of \texttt{StoFedDelAv}}
\vspace{-1mm}
\label{sec:conv}
\noindent In this section, we explore the optimality gap between the model chosen at the latest global aggregation $K$ and the optimal model. We then obtain the optimal model combiner weight. \noindent\textit{All the proofs can be found in our online technical report~\cite{SFDATechReport}}.

\noindent\textbf{Definition 1.} \textit{The local data variability of device $i$ is measured via $\Theta_i\geq0,\forall i$, satisfying$\|\nabla f_i(\textbf x_1,y_1;\textbf{w})-\nabla f_i(\textbf x_2,y_2;\textbf{w})\| \leq\Theta_{i}\|\textbf x_1-\textbf x_2\|, ~\forall (\textbf x_1,y_1),(\textbf x_2,y_2)\in\mathcal{D}_i.$}

\vspace{1mm}
\noindent\textbf{Definition 2.} \textit{For $k\in\{0,...,K-1\}$, the centralized GD during $t\in\{k\tau-\Delta+1,...,(k+1)\tau-\Delta\}$ is defined as $\textbf{c}_k(t)=\textbf{c}_k(t-1)-\eta\nabla F(\textbf{c}_k(t-1))$} \textit{initialized such that $\textbf{c}_k(k\tau-\Delta)=\textbf{w}(k\tau-\Delta)$.} 

We now characterize the variance of SGD noise in~\eqref{eq:SGDwithNoise}:

\noindent{\bf Lemma 1.} {\it Using Definition 1, the SGD noise at each local iteration $t$ at each device $i$ can be upper bounded as follows:}

\vspace{-4.5mm}
{\small
\begin{equation}
	\E{[\|\nu_i(t)\|^2]} \leq \left(1-{|\mathcal{D}_i(t)|}/{N_i}\right){2\left(\Theta_i S_i\right)^2}/{|\mathcal{D}_i(t)|},
\end{equation}}
\vspace{-5mm}

\noindent where $S_i^2$ is the sample variance of data at device $i$.






Since the the minibatch size (i.e, $|\mathcal{D}_i(t)|$, $\forall i$ in the above definition) is fixed during each local training interval and only varies across global aggregations, with some abuse of notation, we replace $t$ with $k$ in the above definition and express the SGD noise during period $k$, using Jensen's inequality, as

\vspace{-2mm}
{\small
\begin{equation}\label{eq:boundedNoiseDef}
	\hspace{-2mm}\E{[\|\nu_i(k)\|]} \leq \Theta_i S_i\sqrt{2\left(1-{n_{i}(k)}/{N_i}\right)/{n_{i}(k)}}.
\end{equation}}

In Theorem $1$, we bound the loss gap, i.e., $F(\textbf{w}^{K})-F(\textbf{w}^\star)$:\\\\
\vspace{-8mm}

\noindent\textbf{Theorem 1.} \textit{If $\eta<\frac{2}{\beta}$, then under Assumption 1, we have}
\vspace{-4mm}

{\footnotesize
\begin{align} 
        &\hspace{-2mm}F(\textbf{w}^{K})-F(\textbf{w}^\star)\leq \frac{1}{2\eta\phi T}+\sqrt{\frac{1}{4\eta^2 \phi^2 T^2}+\frac{L\Psi(\widehat{\bm{\alpha}})}{\eta\phi T}} + L\Psi(\widehat{\bm{\alpha}}) \hspace{-2mm}\nonumber\\
        &~~~~~~~~~~~~~~~~~~~~~~\triangleq \mathcal{L}(\{n_{i}(k)\}_{i\in\mathcal{I},1\leq k\leq K}),\label{eq:thm_1}\vspace{-4mm}\\[-.9em]
    &\hspace{-3mm}\Psi(\widehat{\bm{\alpha}})\triangleq\sum_{k=1}^{K}\psi(\alpha(k),k),\\
    &\hspace{-3mm}\begin{aligned}\label{eq:psi_def}
       \psi(\alpha(k),&k)=\E[\|\textbf{w}((k+1)\tau-\Delta)-\textbf{c}_k((k+1)\tau-\Delta)\|\\
       &\leq(1-\alpha(k))\epsilon(k)([1+\eta\beta]^{\tau}-1)\\
       &+(1-\alpha(k))h(\tau,k)+\alpha(k) h(\tau-\Delta,k)\\
       &+\alpha(k)\eta\Delta L[1+\eta\beta]^{\tau-\Delta}+\eta\sigma(k)[\tau-\alpha(k)\Delta],
    \end{aligned}\\
    &\hspace{-3mm}h(x,k)\triangleq{((\delta+\sigma(k))}/{\beta})[(1+\eta\beta)^{x}-1]-\eta(\delta+\sigma(k))x\label{eq:h_def},\\
    &\hspace{-3mm}\epsilon(k)\triangleq(1-(1-\alpha(k))^k)\left[2\eta(L+\sigma(k))\left({\tau}/{\alpha(k)}-\Delta\right)\right]\label{eq:eps_def},\\
    &\hspace{-3mm}\sigma(k) \triangleq\hspace{-.2mm} \sum_i{\hspace{-.2mm}\rho_i{E[\|\nu_i(k)\|}}=\hspace{-.2mm}\sum_i{\rho_i S_{i}\Theta_{i}\sqrt{2\left(N_i\hspace{-.2mm}-\hspace{-.2mm}n_i(k)\right)\hspace{-.2mm}/\hspace{-.2mm}\left(N_i n_i(k)\right)}}.\hspace{-5mm}
\end{align}}
 \vspace{-4mm}
 
\vspace{-1.5mm}
The optimality gap in \eqref{eq:thm_1} decreases as $T$ increases. More explicitly, as $T,K\rightarrow\infty$, $F(\textbf{w}^{K})-F(\textbf{w}^\star)$ is determined exclusively by $\Psi(\widehat{\bm{\alpha}})$ terms in \eqref{eq:thm_1}.
Critical to the understanding of the behavior of the optimality gap defined in \eqref{eq:thm_1} is $\Psi(\widehat{\bm{\alpha}})$, which comprises terms $\psi(\alpha(k),k)$ given by \eqref{eq:psi_def}. $\Psi(\widehat{\bm{\alpha}})$ ultimately defines the discrepancy between the global model and the theoretical centralized model on one aggregation period. 

It is important to note that the last term of \eqref{eq:h_def} (i.e. the term with a negative sign) is decreasing with respect to (w.r.t.) gradient dissimilarity and noise. In most contexts, however, this term is counteracted by the rest of the terms in $\psi(\alpha(k),k)$ that are increasing w.r.t. SGD noise and gradient dissimilarity.

Crucial to the minimization of \eqref{eq:thm_1} is the proper choice of $\alpha(k)$. Although the behavior of the expression in \eqref{eq:psi_def} is non-trivial to analyze,
we experimentally observe in Fig. $2$(a) (Sec.~\ref{sec:exper_results}) that $\psi(\alpha(k),k)$ is convex as a function of $\alpha(k)\in(0,1]$, implying $[F(\textbf{w}^{K})-F(\textbf{w}^{\star})]\propto\sqrt{\Psi(\widehat{\bm{\alpha}})}+\Psi(\widehat{\bm{\alpha}})$ can be minimized by minimizing each $\psi(\alpha(k),k)$ since $\psi(\alpha(k),k)$'s are independent according to~\eqref{eq:psi_def}. In particular, each $\psi(\alpha(k),k)$ is the solution to the optimization problem
  
\vspace{-2mm}
{\small
\begin{equation}
    \alpha^\star(k)=\argmin_{\alpha(k)\in(0,1)} \psi(\alpha(k),k),
\end{equation}
}

\vspace{-2mm}
\noindent where $\psi(\alpha(k),k)$ is given by~\eqref{eq:psi_def}. Since the closed-form solution of the above problem is non-trivial, this problem can be solved using numerical methods given the bounded range of $\alpha(k)$. Nevertheless, given~\eqref{eq:psi_def} optimizing over $\psi(\alpha(\infty),\infty)$ would give us the following closed-form solution:
\vspace{-4mm}

{\footnotesize
\begin{equation}\label{eq:opt_alpha}
    \begin{aligned}
       &\hspace{-7.5mm}\alpha^\star(\infty)=\min{\left(1,\sqrt{{2\eta\tau(L+\sigma(\infty))[(1+\eta\beta)^{\tau}-1]}/{A}}\right)}\\
       A=&2\eta\Delta(L+\sigma(\infty))[(1+\eta\beta)^{\tau}-1]+\eta\Delta L(1+\eta\beta)^{\tau-\Delta}\\
         &\hspace{-4mm}-({(\delta+\sigma(\infty))}/{\beta})(1+\eta\beta)^{\tau-\Delta}[(1+\eta\beta)^{\Delta}-1]+\eta\delta\Delta.
    \end{aligned}\hspace{1mm}
\end{equation}}
\vspace{-3mm}

\noindent In practice, to avoid a numerical method, one can use \eqref{eq:opt_alpha} for each $\alpha^\star(k)$ with using $\sigma(k)$ instead of $\sigma(\infty)$ in \eqref{eq:opt_alpha}.

\vspace{-1mm}
\section{Network Optimization Problem}\label{sec:opt_prob}
\noindent In this section, we first formulate a problem to jointly minimize energy, delay, and model loss in Sec. \ref{ssec:prob_form}. We then rework the problem and solve it in Sec. \ref{ssec:GP_form}.

\vspace{-1.5mm}
\subsection{Problem Formulation}\label{ssec:prob_form}
For period $k$, let $E^{\mathsf{Cmp}}(k)$ be the energy required to compute the gradient over a minibatch of data, $E^{\mathsf{Tx}}(k)$ be the energy required for model transmission, $T^{\mathsf{Cmp}}(k)$ be the computation time, $T^{\mathsf{Tx}}(k)$ be the model transmission time, $Q$ be the number of bits per model, $p_i(k)$ be the transmit power of device $i$, and $R_i(k)$ be the data rate between device $i$ and the BS.

We formulate the following problem to optimize a trade-off between energy consumption, delay, and model performance:
\vspace{-4mm}

{\footnotesize
\begin{align}
    &\hspace{-1mm}\bm{\mathcal{P}}:\hspace{-2mm}\min_{\{\bm{n}(k)\}_{k=1}^{K}}  \sum_{k=1}^{K} \hspace{-.5mm}\Big[c_1\big[E^{\mathsf{Cmp}}(k)+E^{\mathsf{Tx}}(k)\big] \hspace{-.5mm}+ \hspace{-.5mm}c_2 \big[T^{\mathsf{Cmp}}(k)+T^{\mathsf{Tx}}(k)\big]\Big]\hspace{-2mm} \nonumber\\ 
        &~~~~~~~~~~~~~~~~~~~+ c_3 \mathcal{L}(\{n_{i}(k)\}_{i\in\mathcal{I},1\leq k\leq K})\\[-.4em]
    &\hspace{-1mm} \textrm{s.t.}\nonumber\\
    &\hspace{-1mm}\mathbf{(C1)}\ E^{\mathsf{Cmp}}(k)=\sum_{i\in\mathcal{I}}E_{i}^{\mathsf{Cmp}}(k),\nonumber\\
    &\hspace{-2mm}\mathbf{(C2)}\ E^{\mathsf{Tx}}(k)=\sum_{i\in\mathcal{I}}E_{i}^{\mathsf{Tx}}(k),\vspace{-1mm}\nonumber\\[-0.5em]
    \vspace{-3mm}&\hspace{-2mm}\mathbf{(C3)}\ \sum_{k=1}^{K}E_{i}^{\mathsf{Cmp}}(k)+E_{i}^{\mathsf{Tx}}(k)\leq E_{i}^{\mathsf{Batt}},~\forall i\in\mathcal{I},\nonumber\\
    &\hspace{-2mm}\mathbf{(C4)}\ E_{i}^{\mathsf{Cmp}}(k)= {\gamma_i}d_{i}\tau n_{i}(k)\varrho_{i}^{2}/2,~\forall i\in\mathcal{I,}\nonumber\\
    &\hspace{-2mm}\mathbf{(C5)}\ E_{i}^{\mathsf{Tx}}(k) = p_i(k){Q}/{R_i(k)},~\forall i\in\mathcal{I}, \nonumber\\
    &\hspace{-2mm}\mathbf{(C6)}\ T^{\mathsf{Cmp}}(k)=\max_{i\in\mathcal{I}}\tau {d_{i}n_{i}(k)}/{\varrho_{i}},\nonumber\\
    &\hspace{-2mm}\mathbf{(C7)}\ T^{\mathsf{Tx}}(k)=\max_{i\in\mathcal{I}}{Q}/{R_{i}(k)},\nonumber\\
    &\hspace{-2mm}\mathbf{(C8)}\ \mathcal{L}(\{\bm{n}(k)\}_{k=1}^{K})=F(\textbf{w}^{K})-F(\textbf{w}^\star) ~~(\textrm{see } \eqref{eq:thm_1}),\nonumber\\
    &\hspace{-2mm}\mathbf{(C9)}\ 0\leq n_{i}(k) \leq N_{i}, ~\forall i\in\mathcal{I},\nonumber
\end{align}}
\vspace{-7mm}

\noindent where $\bm{n}(k)=\{{n}_i(k) \}_{i\in\mathcal{I}}$ is the collection of minibatch sizes of the devices over the training interval, and constants $c_1,c_2,c_3\geq 0$ weigh the importance of the objective terms.

Constraints $\mathbf{C1}$ and $\mathbf{C2}$ are, respectively, the total computation and transmission energy consumption during each global aggregation. $\mathbf{C3}$ limits the amount of energy device $i$ can consume over $K$ according to its battery $E_{i}^{\mathsf{Batt}}$. $\mathbf{C4}$ constrains the computation energy of $i$, where $\gamma_i$ is its effective CPU capacitance, $d_i$ is the number of CPU cycles needed to process one datapoint, and $\varrho_{i}$ is the CPU clocking frequency~\cite{wang2019adaptive,tran2019federated}. $\mathbf{C5}$ represents the energy needed for transmission, and constraints $\mathbf{C6}$ and $\mathbf{C7}$ are the computation and transmission time, respectively, for the network. $\mathbf{C8}$ constrains the loss gap to its upper bound, and constraint $\mathbf{C9}$ ensures $\bm{\mathcal{P}}$'s feasibility.

\vspace{-2mm}
\subsection{Geometric Programming-based Optimization}\label{ssec:GP_form}
\vspace{-1.2mm}
Problem $\bm{\mathcal{P}}$ is non-convex, particularly due to the behavior of $\mathcal{L}$ in the objective function. However, by fixing the value of $\alpha(k)$, the problem reduces to a signomial programming (SP) problem \cite{chiangGP}. While this is still NP-hard in general, the resulting SP can be solved via the method of posynomial condensation and penalty functions \cite{9301243}. We thus transform $\bm{\mathcal{P}}$ into an iterative problem in which at each iteration $\ell$, a convex problem is obtained via logarithmic change of optimization variables (c.o.v.), the solution of which is used to determine the value of $\widehat{\bm{\alpha}}$ using \eqref{eq:opt_alpha}. In particular, we write the problem as an optimization problem with a \textit{posynomial} objective function subject to equality on \textit{monomials} and inequality on \textit{posynomails}, which admits the format of geometric programming (GP)~\cite{chiangGP}. 
As a result, at each iteration $\ell$, we aim to find the solution to the following optimization problem, which can undergo a logarithmic c.o.v. and be reduced to a convex problem:
\vspace{-1mm}

\vspace{-.5mm}
{\footnotesize
\begin{table*}[tbp]
\vspace{1.4mm}
\begin{minipage}{0.99\textwidth}
    \begin{equation}\label{eq:f_hat_1}
        f_1(\bm{y},\widehat{\bm{\alpha}})=\sum_{k=1}^{K}\psi(\alpha(k),k)\rightarrow f_1(\bm{y},\widehat{\bm{\alpha}})\geq\widehat{f}_{1}(\bm{y},\widehat{\bm{\alpha}};\ell)\triangleq\prod_{k=1}^{K}\left(\frac{\psi(\alpha(k),k)f_{1}(\bm{y},\widehat{\bm{\alpha}})^{[\ell-1]}}{\psi(\alpha(k),k)^{[\ell-1]}}\right)^{\frac{\psi(\alpha(k),k)^{[\ell-1]}}{f_{1}(\bm{y},\widehat{\bm{\alpha}})^{[\ell-1]}}}
    \end{equation}
    \begin{equation}\label{eq:f_hat_2}
        \begin{aligned}
            f_2(\bm{y},k)=&\underbrace{\alpha(k)\mathsf{B}_{4}(k)\epsilon(k)\mathsf{B}_{1}}_{q_{2,1}}+\underbrace{\mathsf{B}_{4}(k)\mathsf{h}_{1}(k)}_{q_{2,2}}+\underbrace{\alpha(k)\mathsf{h}_{2}(k)}_{q_{2,3}}+\underbrace{\alpha(k)\eta\Delta L\mathsf{B}_{5}}_{q_{2,4}}+\underbrace{\eta\sigma(k)\mathsf{B}_{6}(k)}_{q_{2,5}}\rightarrow\\\vspace{-2.5mm}
            &f_2(\bm{y},k)\geq\widehat{f}_2(\bm{y},k;\ell)\triangleq\prod_{j=1}^{5}\left(\frac{q_{2,j}f_2(\bm{y},k)^{[\ell-1]}}{q_{2,j}^{[\ell-1]}}\right)^{\frac{q_{2,j}^{[\ell-1]}}{f_2(\bm{y},k)^{[\ell-1]}}}
        \end{aligned}
        \vspace{-3mm}
    \end{equation}
    \vspace{-5mm}
    \begin{equation}\label{eq:f_hat_3}
        f_3(\bm{y},x,k,i)=\underbrace{1}_{q_{3,1}}+\underbrace{\mathsf{h}_i^{-1}(k)\eta\delta x}_{q_{3,2}} +\underbrace{\mathsf{h}_i^{-1}(k)\eta\sigma(k)x}_{q_{3,3}}\rightarrow
        f_3(\bm{y},x,k,i)\geq\widehat{f}_{3}(\bm{y},x,k,i;\ell)\triangleq\prod_{j=1}^{3}\left(\frac{q_{3,j}f_{3}(\bm{y},x,k,i)^{[\ell-1]})}{q_{3,j}^{[\ell-1]}}\right)^{\frac{q_{3,j}^{[\ell-1]}}{f_{3}(\bm{y},x,k,i)^{[\ell-1]}}}
    \end{equation}
    \vspace{-6mm}
    \begin{equation}\label{eq:f_hat_4}
        f_{4}(\bm{y},x,k,i)=\underbrace{\mathsf{h}_{i}^{-1}(k)\mathsf{B}_i\delta\beta^{-1}}_{q_{4,1}}+\underbrace{\mathsf{h}_i^{-1}(k)\mathsf{B}_i\sigma(k)\beta^{-1}}_{q_{4,2}}\rightarrow f_{4}(\bm{y},x,k,i)\geq\widehat{f}_{4}(\bm{y},x,k,i;\ell)\triangleq\prod_{j=1}^{2}\left(\frac{q_{4,j}f_{4}(\bm{y},x,k,i)^{[\ell-1]}}{q_{4,j}^{[\ell-1]}}\right)^{\frac{q_{4,j}^{[\ell-1]}}{f_{4}(\bm{y},x,k,i)^{[\ell-1]}}}
    \end{equation} 
    \vspace{-5mm}
    \begin{equation}\label{eq:f_hat_5}
            f_{5}(\bm{y},k)=\underbrace{1}_{q_{5,1}}+\underbrace{\epsilon^{-1}(k)\mathsf{B}_{3}(k)2\eta L\Delta}_{q_{5,2}}+\underbrace{\epsilon^{-1}(k)\mathsf{B}_{3}(k)2\eta\sigma(k)\Delta}_{q_{5,3}}\rightarrow f_{5}(\bm{y},k)\geq\widehat{f}_{5}(\bm{y},k;\ell)\triangleq\prod_{j=1}^{3}\left(\frac{q_{5,j}f_{5}(\bm{y},k)^{[\ell-1]}}{q_{5,j}^{[\ell-1]}}\right)^{\frac{q_{5,j}^{[\ell-1]}}{f_{5}(\bm{y},k)^{[\ell-1]}}}
    \end{equation} 
    \vspace{-6mm}
    \begin{equation}\label{eq:f_hat_6}
        f_{6}(\bm{y},k)=\underbrace{\epsilon^{-1}(k)\mathsf{B}_{3}(k)2\eta L\tau\alpha(k)^{-1}}_{q_{6,1}}+\underbrace{\epsilon^{-1}(k)\mathsf{B}_{3}(k)2\eta\sigma(k)\tau\alpha(k)^{-1}}_{q_{6,2}}\rightarrow f_{6}(\bm{y},k)\geq\widehat{f}_{6}(\bm{y},k;\ell)\triangleq\prod_{j=1}^{2}\left(\frac{q_{6,j}f_{6}(\bm{y},k)^{[\ell-1]}}{q_{6,j}^{[\ell-1]}}\right)^{\frac{q_{6,j}^{[\ell-1]}}{f_{6}(\bm{y},k)^{[\ell-1]}}}
    \end{equation} 
    \vspace{-4mm}
    \begin{equation}\label{eq:f_hat_7}
        f_{7}(\bm{y},k)=\sum_{j\in\mathcal{I}}\rho_{j}S_{j}\Theta_{j}\sqrt{2}P_{j}(k)\rightarrow f_{7}(\bm{y},k)\geq\widehat{f}_{7}(\bm{y},k;\ell)\triangleq\prod_{j\in\mathcal{I}}\left(\frac{(\rho_{j}S_{j}\Theta_{j}\sqrt{2}P_{j}(k))f_{7}(\bm{y},k)^{[\ell-1]}}{\left\{\rho_{j}S_{j}\Theta_{j}\sqrt{2}P_{j}(k)\right\}^{[\ell-1]}}\right)^{\frac{\left\{\rho_{j}S_{j}\Theta_{j}\sqrt{2}P_{j}(k)\right\}^{[\ell-1]}}{f_{7}(\bm{y},k)^{[\ell-1]}}}
    \end{equation}
    \vspace{-4mm}
    \begin{equation}\label{eq:f_hat_8}
        f_{8}(\bm{y},k,i)=\underbrace{P_i^2(k)n_{i}(k)}_{q_{8,1}}+\underbrace{n_{i}(k)N_i^{-1}}_{q_{8,2}}\rightarrow f_{8}(\bm{y},k,i)\geq\widehat{f}_{8}(\bm{y},k,i;\ell)\triangleq\prod_{j=1}^{2}\left(\frac{q_{8,j}f_{8}(\bm{y},k,i)^{[\ell-1]}}{q_{8,j}^{[\ell-1]}}\right)^{\frac{q_{8,j}^{[\ell-1]}}{f_{8}(\bm{y},k,i)^{[\ell-1]}}}
    \end{equation}
\end{minipage}
\hrule
\vspace{-6mm}
\end{table*}}

\vspace{-4mm} 
{\footnotesize 
\begin{align}
    &\hspace{-.55mm}\widehat{\bm{\mathcal{P}}}:~\min_{\bm{y}}
        \sum_{k=1}^{K}\Big[ c_1\big[E^{\mathsf{Cmp}}(k)+E^{\mathsf{Tx}}(k)\big]+c_2 \big[T^{\mathsf{Cmp}}(k)+T^{\mathsf{Tx}}(k)\big]\Big]\nonumber\\
    &\hspace{-.55mm}~~~~~~~+c_3\mathcal{L}(\{\bm{n}(k)\}_{k=1}^{K})\\
    &\hspace{-.55mm}~~~~~~~+w_{1}s_{1}+\sum_{k=1}^{K}\left[\sum_{j=2}^{4}w_{j}(k)s_{j}(k)+\sum_{i\in\mathcal{I}}w_{5}(k,i)s_{5}(k,i)\right]\hspace{-5 mm}\nonumber\\[-.5em] 
    &\hspace{-.55mm} \textrm{s.t.}\nonumber\\[-.3em]
    &\hspace{-.55mm}\mathbf{(\widehat{C}1)}\ \sum_{i\smaller{\in}\mathcal{I}}E_{i}^{\mathsf{Cmp}}(k)/{E^{\mathsf{Cmp}}(k)}\leq 1\nonumber\\[-.2em]
    &\hspace{-.55mm}\mathbf{(\widehat{C}2)}\ \sum_{i\smaller{\in}\mathcal{I}}E_{i}^{\mathsf{Tx}}(k)/{E^{\mathsf{Tx}}(k)}\leq 1\nonumber\\
    &\hspace{-.55mm}\mathbf{(\widehat{C}3)}\ \sum\nolimits_{k=1}^{K}\left(E_{i}^{\mathsf{Cmp}}(k)+E_{i}^{\mathsf{Tx}}(k)\right)/{E_i^{\mathsf{Batt}}}\leq 1, \forall i\in\mathcal{I}\nonumber\\
    &\hspace{-.55mm}\mathbf{(\widehat{C}4)}\ {\gamma_i}d_{i}\tau n_{i}(k)\varrho_{i}^{2}/{(2E_{i}^{\mathsf{Cmp}}(k))} = 1,~\forall i\in\mathcal{I}\nonumber\\\nonumber\\[4ex]
    &\hspace{-.55mm}\mathbf{(\widehat{C}5)}\ {p_{i}Q}/{({E_{i}^{\mathsf{Tx}}(k)}R_{i})} = 1,~\forall i\in\mathcal{I}\nonumber\\  
    &\hspace{-.55mm}\mathbf{(\widehat{C}6)}\ {\tau d_{i}n_{i}(k)}/{(T^{\mathsf{Cmp}}(k)\varrho_{i})} \leq 1,\forall i\in \mathcal{I}\nonumber\\
    &\hspace{-.55mm}\mathbf{(\widehat{C}7)}\ {Q}/{(T^{\mathsf{Tx}}(k)R_{i})} \leq 1,\forall i\in \mathcal{I}\nonumber\\
    &\hspace{-.55mm}\mathbf{(\widehat{C}8.1)}\ \mathcal{L}^{-1}\left[m_1+P_{1}+L \Psi(\widehat{\bm{\alpha}})\right]\leq 1\nonumber\\
    &\hspace{-.55mm}\mathbf{(\widehat{C}8.2)}\ (m_1 2\eta\phi T)^{-1}\leq 1\nonumber\\
    &\hspace{-.55mm}\mathbf{(\widehat{C}8.3)}\ P_1^{-2}(m_2+m_3\Psi(\widehat{\bm{\alpha}}))\leq 1\nonumber\\
    &\hspace{-.55mm}\mathbf{(\widehat{C}8.4)}\ m_2^{-1}({1}/{(2\eta\phi T}))^2=1\nonumber\\
    &\hspace{-.55mm}\mathbf{(\widehat{C}8.5)}\ {L m_3^{-1}}/{(\eta\phi T)}=1\nonumber\\
    &\hspace{-.55mm}\mathbf{(\widehat{C}8.6)}\ \Psi^{-1}(\widehat{\bm{\alpha}})\sum_{k=1}^{K}\psi(\alpha(k),k)\leq 1\nonumber\\
    &\hspace{-.55mm}\mathbf{(\widehat{C}8.7)}\ {s_1^{-1}\Psi(\widehat{\bm{\alpha}})}/{\widehat{f}_1(\bm{y},\widehat{\bm{\alpha}};\ell)}\leq 1\nonumber\\
    &\hspace{-.55mm}\mathbf{(\widehat{C}8.8)}\ \psi^{-1}(k)[\alpha(k)\mathsf{B}_{4}(k)\epsilon(k)\mathsf{B}_{1}+\mathsf{B}_{4}(k)\mathsf{h}_{1}(k)\nonumber\nonumber\\
    &\hspace{10mm}~~~~~~~~~~+\alpha(k) \mathsf{h}_{2}(k)+\alpha(k)\eta\Delta L\mathsf{B}_{5}+\eta\sigma(k)\mathsf{B}_{6}(k)]\leq 1\nonumber\\
    &\hspace{-.55mm}\mathbf{(\widehat{C}8.9)}\ {s_2^{-1}(k)\psi(k)}/{\widehat{f}_2(\bm{y},k;\ell)}\leq 1\nonumber\\
    &\hspace{-.55mm}\mathbf{(\widehat{C}8.10)}\ ({\mathsf{h}_{1}^{-1}(k)\mathsf{B}_{1}\delta\beta^{-1}+\mathsf{h}_{1}^{-1}(k)\mathsf{B}_{1}\sigma(k)\beta^{-1}})/{\widehat{f}_3(\bm{y},\tau,k,1;\ell)}\leq 1\nonumber\\
    &\hspace{-.55mm}\mathbf{(\widehat{C}8.11)}\ \frac{s_{3}^{-1}(k)\left[1+\mathsf{h}_{1}^{-1}(k)\eta\delta\tau +\mathsf{h}_{1}^{-1}(k)\eta\sigma(k)\tau\right]}{\widehat{f}_4(\bm{y},\tau,k,1;\ell)}\leq 1\nonumber\\
    &\hspace{-.55mm}\mathbf{(\widehat{C}8.12)}\ \frac{\mathsf{h}_{2}^{-1}(k)\mathsf{B}_{2}\delta\beta^{-1}+\mathsf{h}_{2}^{-1}(k)\mathsf{B}_{2}\sigma(k)\beta^{-1}}{\widehat{f}_3(\bm{y},\tau-\Delta,k,2;\ell)}\leq 1\nonumber\\
    &\hspace{-.55mm}\mathbf{(\widehat{C}8.13)}\ \frac{s_{4}^{-1}(k)\left[1+\mathsf{h}_{2}^{-1}(k)\eta\delta\mathsf{B}_{7} +\mathsf{h}_{2}^{-1}(k)\eta\sigma(k)\mathsf{B}_{7}\right]}{\widehat{f}_4(\bm{y},\tau-\Delta,k,2;\ell)}\leq 1\nonumber\\
    &\hspace{-.55mm}\mathbf{(\widehat{C}8.14)}\ {\epsilon(k)^{-1}\mathsf{B}_{3}(k)2\eta (L+\sigma(k))\alpha(k)^{-1}}/{\widehat{f}_5(\bm{y},k;\ell)}\leq 1\nonumber\\
    &\hspace{-.55mm}\mathbf{(\widehat{C}8.15)}\ \left({s_{5}^{-1}(k)\left[1+\epsilon(k)^{-1}\mathsf{B}_{3}(k)2\eta (L+\sigma(k))\Delta\right]}\right)/{\widehat{f}_{6}(\bm{y},k;\ell)}\leq 1\nonumber\\
    &\hspace{-.55mm}\mathbf{(\widehat{C}8.16)}\ \sum_{i\in\mathcal{I}}\rho_{i}S_{i}\Theta_{i}\sqrt{2}P_{i}(k)/{\sigma(k)}\leq 1\nonumber\\
    &\hspace{-.55mm}\mathbf{(\widehat{C}8.17)}\ {s_{6}^{-1}(k)\sigma(k)}/{\widehat{f}_{7}(\bm{y},k;\ell)}\leq 1\nonumber\\
    &\hspace{-.55mm}\mathbf{(\widehat{C}8.18)}\ P_i^2(k)n_{i}(k)+n_{i}(k)N_i^{-1}\leq 1,~\forall i\in\mathcal{I}  \nonumber\\
    &\hspace{-.55mm}\mathbf{(\widehat{C}8.19)}\ {s_{7}^{-1}(k,i)}/{\widehat{f}_{8}(\bm{y},k;\ell)}\leq 1,~\forall i\in\mathcal{I}\nonumber\\
    &\hspace{-.55mm}\mathbf{(\widehat{C}9)}\ 0\leq n_{i}(k) \leq N_{i}, ~\forall i\in\mathcal{I}\nonumber\\
    &\hspace{-.55mm}\mathbf{(\widehat{C}10)}\ \left\{s_1,\left\{\bm{s}_{j}(k)\right\}_{2\leq j\leq 6, 1\leq k\leq K},
        \left\{\bm{s}_{7}(k,i)\right\}_{i\in\mathcal{I},1\leq k\leq K}\right\}\geq 1\nonumber
    \end{align}} 
    \vspace{-9mm}    

    {\footnotesize
    \begin{align}
    &\hspace{-.1mm}\textrm{\textbf{Variables}:}~\hspace{-.4mm}{\footnotesize \bm{y}\hspace{-.5mm}\triangleq\hspace{-.5mm}\bigg\{ \hspace{-.4mm}P_{1},\hspace{-.2mm}\Psi(\widehat{\bm{\alpha}}),\hspace{-.2mm}\Big\{\hspace{-.2mm}T^{\mathsf{Cmp}}(k),\hspace{-.2mm}T^{\mathsf{Tx}}(k),\hspace{-.2mm}E^{\mathsf{Cmp}}(k),\hspace{-.2mm}E^{\mathsf{Tx}}(k)}\hspace{-.4mm}\Big\}_{k=1}^{K},\nonumber\\
        &{\footnotesize\hspace{5mm}  \Big\{\{\bm{n}(k)\},\sigma(k),\mathsf{h}_{1}(k),\mathsf{h}_{2}(k),\epsilon(k),\psi(k),\{P_{i}(k)\}_{i\in\mathcal{I}}\Big\}_{k=1}^{K},\nonumber}\\
        &{\footnotesize \hspace{5mm}s_1,\ \left\{\bm{s}_{j}(k)\right\}_{2\leq j\leq 6, 1\leq k\leq K},
        \left\{\bm{s}_{7}(k,i)\right\}_{i\in\mathcal{I},1\leq k\leq K}}\bigg\},
        \nonumber
    \end{align}}
\vspace{-5mm}

\noindent where $\mathsf{h}_{1}(k)=h(\tau,k)$, $\mathsf{h}_{2}(k)=h(\tau-\Delta,k)$,  $\mathsf{B}_1=(1+\eta\beta)^{\tau}-1$, $\mathsf{B}_2=(1+\eta\beta)^{\tau-\Delta}-1$, $\mathsf{B}_{3}(k)=(1-(1-\alpha(k))^k)$, $\mathsf{B}_{4}(k)=(1-\alpha(k))$, $\mathsf{B}_{5}=(1+\eta\beta)^{\tau-\Delta}$, $\mathsf{B}_{6}(k)=\tau-\alpha(k)\Delta$, and $\mathsf{B}_7=(\tau-\Delta)$. $\mathsf{B}_{j}\geq 0,\forall j$. The $\{s_{j}\}$ terms are added to expand the solution space of each iteration that will be forced to converge to $1$ when the problem is solved using the penalty terms (i.e., ${w}_j \gg 1$, $\forall j$). The terms $\widehat{f}_x(\bm{y},...;\ell)$ approximate posynomial denominators in $\bm{\widehat{\mathcal{P}}}$ as monomials to satisfy the requirements of GP, and are outlined in \eqref{eq:f_hat_1}-\eqref{eq:f_hat_8}. As the iterations progress, these approximations converge towards the value of the posynomial they represent. After convergence, \eqref{eq:opt_alpha} is applied with $\sigma(k)^{[\ell]}$ to update $\widehat{\bm{\alpha}}$ and $\mathsf{B}_{\{3,4,5\}}(k)$. A new problem is then solved given the values of these variables, and this \textit{alternative} process is continued upon convergence.

In $\widehat{\bm{\mathcal{P}}}$, constraints $\mathbf{\widehat{C}1}$-$\mathbf{\widehat{C}5}$ are naturally obtained from problem $\bm{\mathcal{P}}$'s $\mathbf{C}1$-$\mathbf{C}5$ into ones which fit a geometric programming (GP) paradigm. Constraints $\mathbf{\widehat{C}6}$ and $\mathbf{\widehat{C}7}$ stem from the fact that dividing $\bm{\mathcal{P}}$'s $\mathbf{C6}$-$\mathbf{C7}$ computation/transmission times by the maximum computation/transmission time across the network will upper-bound the constraint to $1$. $\bm{\widehat{\mathcal{P}}}$'s constraints $\mathbf{\widehat{C}8.\{1,2,...,19\}}$ develop the transformation of the loss gap of \eqref{eq:thm_1} into a series of constraints in the form of inequalities on posynomials, which is desired in GP programming to have convergence to a Karush–Kuhn–Tucker condition of $\bm{\mathcal{P}}$~\cite{chiangGP}.

\vspace{-1.3mm}
\section{Experimental Results} \label{sec:exper_results}
\vspace{-1mm}
\noindent\textbf{Experimental Setup:} We consider an edge network of $N=5$ devices realized according to the parameters described in Table~\ref{tab:network_setup}. Sets of $N$ parameters are uniformly generated then sorted for $\gamma_i$ and $d_i$ (i.e. $\bm{\gamma}=\{\gamma_{1},...,\gamma_{5}\}$ and $\bm{d}=\{d_{1},...,d_{5}\}$), such that $\gamma_{1}=\argmin{\{\gamma_{i}\}_{i=1}^5}$, $d_{1}=\argmin{\{d_{i}\}_{i=1}^5}$ and $\gamma_{5}=\argmax{\{\gamma_{i}\}_{i=1}^5}$, $d_{5}=\argmax{\{d_{i}\}_{i=1}^5}$. The first device is modeled using $\gamma_{1}$ and $d_{1}$ for its CPU capacitance and number of CPU cycles per datapoint, respectively, making it the most resource-efficient device for data computation; the second device uses $\gamma_{2}$, $d_{2}$, and so on. CVX is used to solve the convex problem at each iteration of $\widehat{\bm{\mathcal{P}}}$. Each plot in Fig. $2$ shows the average of $20$ randomized network initializations. 

\vspace{-.75mm}
{\footnotesize
\begin{table}[!h]
\caption{Parameter settings for experiments.}
\label{tab:network_setup}
\centering
\begin{tabular}{cc}
\hline
\textbf{Parameter(s)} & \textbf{Value / Range}\\
\hline
$c_1,c_2,c_3$ & $1\times 10^{-4},1\times 10^{-3},2.5\times 10^{6}$\\
$E_{i}^{\mathsf{Batt}} \ \forall i$ & $7.5\times 10^{6} (J)$\\
$n_i(k)$ & $[1, 25]$\\
$\varrho_i$ & $1\times 10^{6}(Hz)$\\
$d_i$ & $600\leq d_i \leq 640$\\
$\gamma_i$ & $[4\times 10^{-12}, 6.5\times 10^{-12}](F)$\\
$p_i,R_i,Q$ & $0.1(W)\forall i,~1.0 ~(Mbps) \forall i~, 16~ (kbits)$\\
$\Theta_i,S_i,\delta$ & $2.0 \forall i\ , \ 0.2 \forall i\ , \ 0.5$\\
$\eta,\beta,L,\phi$ & $0.02, \ 1, \ 25, \ 0.025$\\
$\tau,\Delta,K$ & $20, \ 19, \ 15$\\
$w_{1},w_{\{2,...,6\}}(k), \ w_{7}(k,i)$ &  $100000\ , \ 100000 \ , 1000000$\\

\hline
\end{tabular}\\
\vspace{-5mm}
\end{table}}

{\footnotesize
\begin{figure*}[!t]
\centering
\hfil
\subfloat[\centering{$\psi(k)$ as function of $\sigma(k)$ at different periods $k$.}]{\includegraphics[width=2in]{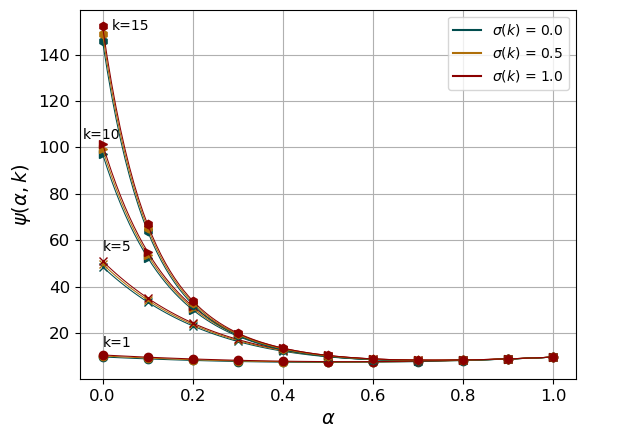}
\label{fig:psi_fn_of_sigma}}
\hfil
\subfloat[\centering{Minibatch size across network over training time.}]{\includegraphics[width=2in]{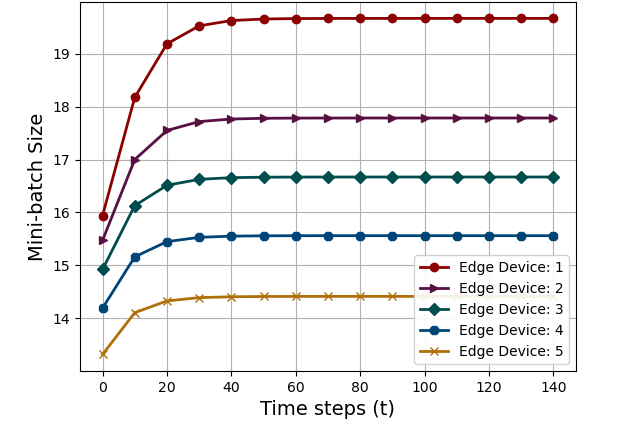}
\label{fig:MB_over_time}}
\hfil
\subfloat[\centering{Average minibatch size across network over cycle $K$ for varying $c_1$.}]{\includegraphics[width=2in]{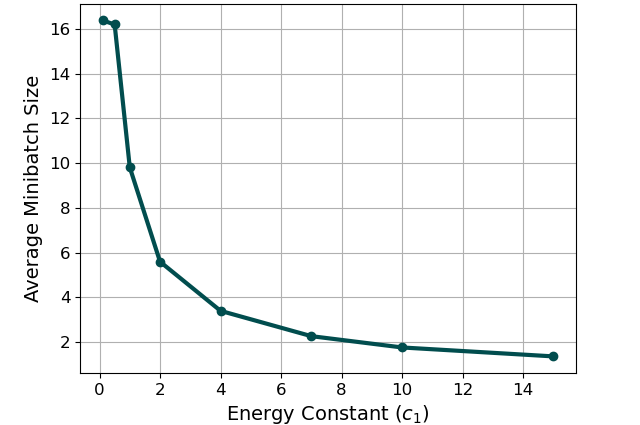}
\label{fig:mb_vs_c_1}}
\hfil
\\
\vspace{-4mm}
\hfil
\subfloat[\centering{Objective value for optimal and fixed $\widehat{\bm{\alpha}}$.}]{\includegraphics[width=2in]{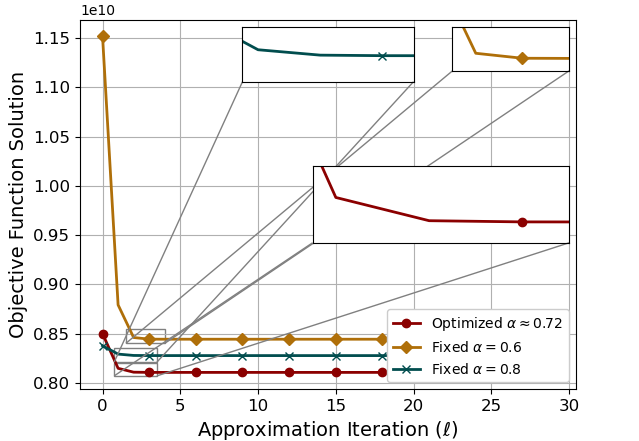}
\label{fig:opt_v_fix_alpha}}
\vspace{-4mm}
\hfil
\subfloat[\centering{$\alpha$ as function of time delay $\Delta$.}]{\includegraphics[width=2in]{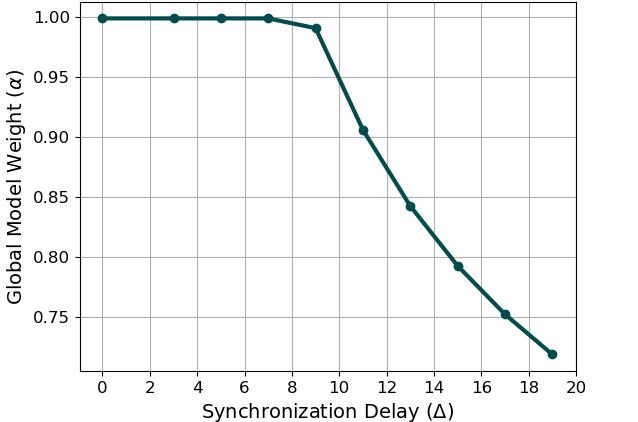}
\label{fig:alpha_fn_of_Del}}
\vspace{-4mm}
\hfil
\subfloat[\centering{Accuracy of logistic regression using MNIST dataset for different fixed $\widehat{\bm{\alpha}}$}]{\includegraphics[width=2in]{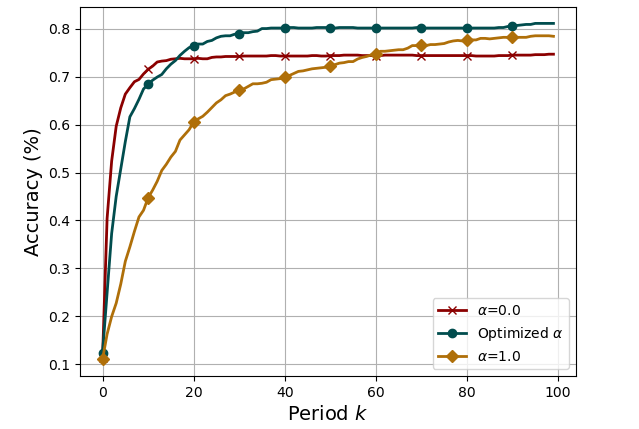}}
\label{fig:empirical_accuracy}
\vspace{4mm}
\caption{Experimental results for our methodology.}
\vspace{-6.5mm}
\label{fig:code_outputs}
\end{figure*}}

\noindent\textbf{Minibatch Optimization:} We first look at minibatch size, which ultimately determines time, energy, and loss across the training interval. Since $\epsilon(k)$ in \eqref{eq:eps_def} becomes more dependent on noise as training progresses due to the term $(1-(1-\alpha(k))^k$, minibatch size should theoretically increase non-linearly over time. This is corroborated in Fig. $2$(b). It can be seen that minibatch size for the devices follows their relative precedence, such that the best edge device, $1$, possesses the largest minibatch, $2$ the second largest, and so on. Better devices show larger differences between their initial minibatch size and their latest, with device $1$ experiencing a nearly $25\%$ increase. This trend reveals that the better edge devices save energy in early training stages for later on when SGD noise is more impactful on the machine learning loss.

\noindent\textbf{Energy and Minibatch:}\label{ssec:energy_and_minibatch}
In Fig $1$(c), we depict average minibatch size across the network while varying the energy constant, $c_1$ in the objective function of $\bm{\mathcal{P}}$. The results show that the precedence assigned to energy and the average minibatch size across the network for the complete training cycle exhibit a steep ramp-down from $c_1\in(0,1)$.

\noindent\textbf{Impact and Behavior of $\bm{\alpha(k)}$:}
By allowing the network to choose $\widehat{\bm{\alpha}}$ per \eqref{eq:opt_alpha}, the value for the objective function of the problem $\widehat{\bm{\mathcal{P}}}$ drops meaningfully, as seen in Fig. $2$(d). We determined numerically that the difference between the calculated optimal value of $\alpha(0)$ and that found using \eqref{eq:opt_alpha} was about $0.16$. Subsequent values of $\alpha(k>0)$ were effectively identical to the numerically optimal value, thus proving the efficacy of the proposed method. It is worth noting that this is feasible for the iterative GP approach, as previous values for $\sigma(k)$ can be used, but in real-time this may not be the case.

$\alpha(k)$ is also heavily dependent on delay as shown in Fig. $2$(e), where the vertical axis represents the average of elements in $\widehat{\bm{\alpha}}$ and the horizontal axis $\Delta$. This shows that the proportionality between $\tau$ and $\Delta$ should be carefully considered when choosing $\widehat{\bm{\alpha}}$. As $\Delta\rightarrow 0$, $\alpha(k)\rightarrow 1$, as is expected in ideal {\tt FedAvg}.

Figure $2$(f) illustrates the impact of $\alpha$ on model accuracy in a real model, trained on the MNIST dataset using $\alpha=0.0,1.0$ and $\alpha_{opt}$. It is readily apparent that optimizing $\alpha$ plays a key role in convergence under the {\tt StoFedDelAv} paradigm.

\vspace{-2.5mm}
\section{Conclusion and Future Work}
\vspace{-2mm}
\noindent We proposed a novel methodology for optimizing federated learning implementations over edge networks while explicitly taking into account device-server communication delay and device computation heterogeneity. The loss optimality gap was considered across a training cycle to characterize the performance of the network. We formulated an optimization problem aiming to find the minibatch size of the devices across the training interval to optimize a trade-off between energy consumption, time required to train the model, and ML model performance. This problem was optimized using an iterative geometric programming-based approach to find the ideal minibatch size for each device across the network. Future works will focus on improving the network and training efficiency, namely distributed device orchestration and delay-aware device sampling. These approaches will enable networks to train models in a more time- and energy-efficient manner. 
\vspace{-1mm}

\bibliographystyle{IEEEtran}
\bibliography{references}

\section{Appendix A}

\subsection{Proof of Lemma 1} 

\noindent\textbf{Lemma 1.}
\textit{For ease of manipulation in bounding equations using the triangular inequality, the noise can be defined as}
\begin{equation}\label{Lem1BoundNoise}
    \E{[\|\nu_i(k)\|]} \leq S_i\Theta_i\sqrt{2}\sqrt{\frac{N_i-n_i(k)}{N_in_i(k)}}
\end{equation}
\noindent\textit{Proof.} We begin by defining the variance of the gradients, $\widehat{\mathcal{S}_i^2}$. With $\bm{\lambda}_i$ and $S_i$ denoting the mean and sample variance of the device's datapoints, respectively, and using Definition 1, we say:

{\small
\begin{align}
    \hspace{-18mm}\widehat{\mathcal{S}_i^2}&=\frac{\sum_{x_1\in \mathcal{D}_i}\|\nabla f_i(x_1,y_1;\textbf{w})-\sum_{x_2\in \mathcal{D}_i}\frac{\nabla f_i(x_2,y_2;\textbf{w})}{N_i})\|^2}{N_i-1}\nonumber\\
    &=\frac{\sum_{x_1\in \mathcal{D}_i}\frac{1}{N_i^2}\|N_i\nabla f_i(x_1,y_1;\textbf{w})-\sum_{x_2\in \mathcal{D}_i}\nabla f_i(x_2,y_2;\textbf{w})\|^2}{N_i-1}\nonumber\\
    &\leq \frac{\sum_{x_1\in \mathcal{D}_i(k)}\frac{N_i-1}{N_i^2}\sum_{x_2\in \mathcal{D}_i}\|\nabla f_i(x_1,y_1;\textbf{w})-\nabla f_i(x_2,y_2;\textbf{w})\|^2}{Z_1-1}\nonumber\\
    &\leq \frac{\sum_{\textbf{x}_1\in \mathcal{D}_i}\frac{(N_i-1)\Theta_i^2}{N_i^2}\sum_{\textbf{x}_2\in \mathcal{D}_i}\|\textbf{x}_1-\textbf{x}_2\|^2}{N_i-1}\nonumber\\
    &\leq\frac{(N_i-1)\Theta_i^2}{N_i^2}\frac{\sum_{x_1\in \mathcal{D}_i}\sum_{x_2\in \mathcal{D}_i}\|\textbf{x}_1-\textbf{x}_2+\bm{\lambda}_i-\bm{\lambda}_i\|^2}{N_i-1}\nonumber\\
    &=\frac{(N_i-1)\Theta_i^2}{N_i^2}\times\nonumber\\
    &\left[\frac{\sum_{x_1\in \mathcal{D}_i}\sum_{x_2\in \mathcal{D}_i}\left[\|\textbf{x}_1-\bm{\lambda}_i\|^2+\|\textbf{x}_2-\bm{\lambda}_i\|^2-2(\textbf{x}_1-\bm{\lambda}_i)^{\mathsf{T}}(\textbf{x}_2-\bm{\lambda}_i)\right]}{N_i-1}\right]\nonumber\\
    &=\frac{(N_i-1)\Theta_i^2}{N_i^2}\frac{N_i\sum_{x_1\in \mathcal{D}_i}\|\textbf{x}_1-\bm{\lambda}_i\|^2+N_i\sum_{x_2\in \mathcal{D}_i}\|\textbf{x}_2-\bm{\lambda}_i\|^2}{N_i-1}\nonumber\\
    &=\frac{2(N_i-1)\Theta_i^2 S_i^2}{N_i}\leq2(\Theta_i S_i)^2,
\end{align}}

\noindent where the first inequality is found using the Cauchy-Schwarz inequality, and the second to last line stems from the fact that $\sum_{x_1\in \mathcal{D}_i}(\textbf{x}_1-\bm{\lambda}_i)=\bm{0}$.

We now look to the variance of the SGD noise itself. As defined in (\ref{eq:boundedNoiseDef}), the variance of the noise for any iteration $\ell$ is
\begin{equation}
    \E[\|\nu_i(t)\|^2]=\left(1-\frac{n_i(t)}{N_i}\right)\frac{\widehat{\mathcal{S}_i^2}}{n_i(t)}.
\end{equation}

\noindent Using the above derivation of $\widehat{\mathcal{S}_i^2}$, we can upper-bound this as

\begin{equation}
    \E[\|\nu_i(t)\|^2]\leq\left(1-\frac{|\mathcal{D}(t)|}{N_i}\right)\frac{2(\Theta_{i}S_i)^2}{|\mathcal{D}(t)|}.
\end{equation}

\noindent Since the minibatch size, $|\mathcal{D}_i(t)|,\forall i$, is fixed during each local training period (i.e. it only varies across global aggregations), with some abuse of notation we replace $t$ with $k$ in the above definition and express the SGD variance during period $k$ as 

\begin{equation}
    \E[\|\nu_i(k)\|^2]\leq\left(1-\frac{n_i(k)}{N_i}\right)\frac{2(\Theta_{i}S_i)^2}{n_i(k)},
\end{equation}

For use in future derivations, we then take the square root of both sides of the equation:

\begin{equation}
    \sqrt{\E[\|\nu_i(k)\|^2]}\leq\Theta_{i}S_i\sqrt{2}\sqrt{\left(1-\frac{n_i(k)}{N_i}\right)\frac{1}{n_i(k)}}.
\end{equation}

\noindent Additionally, by the concavity of a square root function and Jensen's inequality, which states that $\E{[f(X)]}\leq f(\E{[X]})$ for some differentiable, concave function $f$,

\begin{equation}
    \E[\sqrt{\|\nu_i(k)\|^2}] \leq \sqrt{\E[\|\nu_i(k)\|^2]}\leq\Theta_{i}S_i\sqrt{2}\sqrt{\frac{N_i-n_i(k)}{N_i\times n_i(k)}}.
\end{equation}

\noindent Thus the lemma is proven.\hspace{46mm}$\square$

\subsection{Proof of Lemma 2}
\noindent\textbf{Lemma 2.} \textit{Taking the weighted average of Lemma 1 yields a form useful to manipulations necessary in later lemmas, i.e.}
\begin{equation}\label{Lem2SgmaWgtAvg}
    \sigma(k) \triangleq \sum_i{\rho_i\E[\|\nu_i(k)\|]}=\sum_i{\rho_i \Theta_{i}S_i\sqrt{2}\sqrt{\frac{N_i-n_i(k)}{N_i\times n_i(k)}}}
\end{equation}

\subsection{Proof of Lemma 3}
\noindent\textbf{Lemma 3.} 
\begin{equation}\label{Lem43ipschitz}
    \|\nabla F_i(\textbf{w})\|\leq L, \forall i, \forall \textbf{w}
\end{equation}
\textit{Proof.} From the convexity and $L$-Lipschitz conditions, for $\forall\textbf{w}',\textbf{w}$,
\begin{equation}
    \langle\textbf{w}'-\textbf{w},\nabla F_i(\textbf{w})\rangle\leq F_i(\textbf{w}')-F_i(\textbf{w})
\end{equation}
\begin{equation}
    F_i(\textbf{w}')-F_i(\textbf{w})\leq L\|\textbf{w}'-\textbf{w}\|
\end{equation}
Letting $\textbf{w}'=\textbf{w}-\nabla F_i(\textbf{w})$,
\begin{equation}
    \|\nabla F_i(w)\leq L\|
\end{equation}

\subsection{Proof of Lemma 4}
\noindent\textbf{Lemma 4.} \textit{With $\eta<\frac{2}{\beta}$, under Assumption 1,}
\begin{equation}\label{Lem4LocEpsilon}
    \begin{aligned}
    \epsilon_i(k)&\triangleq\|\textbf{w}_i(k\tau-\Delta)-\textbf{w}(k\tau-\Delta)\|\\
    &\leq(1-(1-\alpha(k))^k)[2\eta L\left(\frac{\tau}{\alpha(k)}-\Delta\right)\\
    &+\eta\left(\frac{\tau}{\alpha(k)}-\Delta\right)\sum_j{\rho_j\|\nu_j\|}+\|\nu_i(k)\|]\\
    \end{aligned}
\end{equation}
\textit{Proof.} Using the SGD approximation $g_i$ (which for brevity will have the $\mathcal{D}$ term not included), and letting $\ell=k\tau-r$ and $m=(k+1)\tau-r-\Delta$, we can say
\begin{equation}
    \begin{aligned}
        &\textbf{w}_i((k+1)\tau-\Delta)-\textbf{w}((k+1)\tau-\Delta)\\
        &=\textbf{w}_i((k+1)\tau-\Delta)-\sum_j{\rho_j\textbf{w}_j((k+1)\tau-\Delta)}\\
        &=(1-\alpha(k))[\textbf{w}_i(k\tau-\Delta)-\textbf{w}(k\tau-\Delta)]\\
        &-(1-\alpha(k))\eta\sum_{r=1}^{\Delta}g_i(\textbf{w}_i(\ell))\\
        &+(1-\alpha(k))\eta\sum_j{\rho_j\sum_{r=1}^{\Delta}g_j(\textbf{w}_j(\ell))}\\
        &-(1-\alpha(k))\eta\sum_{r=1}^{\tau-\Delta}{g_i(\textbf{w}_i(m))}\\
        &+(1-\alpha(k))\eta\sum_j{\rho_j\sum_{r=1}^{\tau-\Delta}g_j(\textbf{w}_j(m))}\\\\
        &=(1-\alpha(k))[\textbf{w}_i(k\tau-\Delta)-\textbf{w}(k\tau-\Delta)]\\
        &+(1-\alpha(k))\eta\sum_{r=1}^{\Delta}\left[\sum_j \rho_j g_j(\ell)-g_i(\textbf{w}_i(\ell))\right]\\
        &+\eta\sum_{r=1}^{\tau-\Delta}{\left[\sum_j \rho_j g_j(\textbf{w}_j(m))-g_i(\textbf{w}_i(m))\right]}
    \end{aligned}
\end{equation}
Now expanding the SGD approximations into their gradients and noises,
\begin{equation}
    \begin{aligned}
        &\textbf{w}_i((k+1)\tau-\Delta)-\textbf{w}((k+1)\tau-\Delta)\\
        &=(1-\alpha(k))[\textbf{w}_i(k\tau-\Delta)-\textbf{w}(k\tau-\Delta)]\\
        &+(1-\alpha(k))\eta\sum_{r=1}^{\Delta}[\sum_{j\neq i}\rho_j\nabla F_j(\textbf{w}_j(\ell))\\&+\rho_i\nabla F_i(\textbf{w}_i(\ell))-\nabla F_i(\textbf{w}_i(\ell))\\
        &+\sum_j{\rho_j\nu_j(k)}-\nu_i(k)]\\
        &+\eta\sum_{r=1}^{\tau-\Delta}[\sum_{j\neq i}\rho_j\nabla F_j(\textbf{w}_j(m))\\&+\rho_i\nabla F_i(\textbf{w}_i(m))-\nabla F_i(\textbf{w}_i(m))\\
        &+\sum_j{\rho_j\nu_j(k)}-\nu_i(k)]\\
    \end{aligned}
\end{equation}
Using the triangle inequality and rearranging terms,
\begin{equation}
    \begin{aligned}
        &\|\textbf{w}_i((k+1)\tau-\Delta)-\textbf{w}((k+1)\tau-\Delta)\|\\
        &\leq(1-\alpha(k))\|[\textbf{w}_i(k\tau-\Delta)-\textbf{w}(k\tau-\Delta)]\|\\
        &+(1-\alpha(k))\eta(1-\rho_i)\sum_{r=1}^{\Delta}\|\nabla F_i(\textbf{w}_i(\ell))\|\\
        &+(1-\alpha(k))\eta\sum_{r=1}^{\Delta}\sum_{j\neq i}\rho_j\|\nabla F_j(\textbf{w}_j(\ell))\|\\
        &+(1-\alpha(k))\eta\sum_{r=1}^{\Delta}\left[\sum_j\rho_j\|\nu_j(k)\|+\|\nu_i(k)\|\right]\\
        &+\eta(1-\rho_i)\sum_{r=1}^{\tau-\Delta}\|\nabla F_i(\textbf{w}_i(m))\|\\
        &+\eta\sum_{r=1}^{\tau-\Delta}\sum_{j\neq i}\rho_j \|\nabla F_j(\textbf{w}_j(m))\|\\
        &+\eta\sum_{r=1}^{\tau-\Delta}\left[\sum_j\rho_j\|\nu_j(k)\|+\|\nu_i(k)\|\right]
    \end{aligned}
\end{equation}

Applying Lemma 3 and Assumption 1,
\begin{equation}
    \begin{aligned}
        &\|\textbf{w}_i((k+1)\tau-\Delta)-\textbf{w}((k+1)\tau-\Delta)\|\\
        &\leq(1-\alpha(k))\|[\textbf{w}_i(k\tau-\Delta)-\textbf{w}(k\tau-\Delta)]\|\\
        &+2\eta L(1-\rho_i)(\tau-\alpha(k)\Delta)\\
        &+\eta(\tau-\alpha(k)\Delta)\left[\sum_j\rho_j\|\nu_j(k)\|+\|\nu_i(k)\|\right]
    \end{aligned}
\end{equation}

Recursively unpacking the term until $t=-\Delta$, since $\textbf{w}_i(-\Delta)=\textbf{w}(-\Delta)$,\\\\
\begin{equation}
    \hspace{-5mm}\begin{aligned}
        &\|\textbf{w}_i(k\tau-\Delta)-\textbf{w}(k\tau-\Delta)\|\\
        &\leq(1-\alpha(k))\|\textbf{w}_i(-\Delta)-\textbf{w}(-\Delta)\|\\
        &+(1-(1-\alpha(k))^k)\left[2\eta L\left(\frac{\tau}{\alpha(k)}-\Delta\right)\right]\\
        &+(1-(1-\alpha(k))^k)\left[\eta\left(\frac{\tau}{\alpha(k)}-\Delta\right)\left[\sum_j\rho_j\|\nu_j(k)\|+\|\nu_i(k)\|\right]\right]\\
        &\triangleq\epsilon_i(k)
    \end{aligned}
\end{equation}

\subsection{Proof of Lemma 5}
\noindent\textbf{Lemma 5.} \textit{Taking weighted average of Lemma 4 and applying Lemma 2,}

\begin{equation}\label{Lem5GlobalEpsilon}
    \begin{aligned}
        \epsilon(k)&\triangleq\E\left[\sum_i\rho_i\epsilon_i(k)\right]\\
                  &=(1-(1-\alpha(k))^k)\left[2\eta(L+\sigma(k))\left(\frac{\tau}{\alpha(k)}-\Delta\right)\right]
    \end{aligned}
\end{equation}
\textit{Proof.} First taking the weighted average of all $\epsilon_i(k)$ terms,
\begin{equation}
    \begin{aligned}
        &\sum_i\rho_i\epsilon_i(k)\\
        &=(1-(1-\alpha(k))^k)\{2\eta L(\tau/\alpha(k)-\Delta)\\
        &+\eta(\tau/\alpha(k)-\Delta)[\sum_j\rho_j\|\nu_j(k)\|+\sum_i\rho_i\|\nu_i(k)\|]\}
    \end{aligned}
\end{equation}

Now taking the expectation,
\begin{equation}
    \begin{aligned}
        &\E\left[\sum_i\rho_i\epsilon_i(k)\right]\\
        &=(1-(1-\alpha(k))^k)\{2\eta L(\tau/\alpha(k)-\Delta)\\
        &+\eta(\tau/\alpha(k)-\Delta)[\sum_j\rho_j\E[\|\nu_j(k)\|]\\&+\sum_i\rho_i\E[\|\nu_i(k)\|]]\}\\
        &=(1-(1-\alpha(k))^k)\{2\eta L(\tau/\alpha(k)-\Delta)\\
        &+\eta(\tau/\alpha(k)-\Delta)(2\sigma(k))\}
    \end{aligned}
\end{equation}
Thus proving the lemma after algebraic manipulations.\\

\subsection{Proof of Lemma 6.}
\noindent\textbf{Lemma 6.} \textit{Under Assumption 1, we have}
\begin{equation}\label{Lem6OnePlusEtaBeta}
    \|[\textbf{w}_1-\eta\nabla F(\textbf{w}_1)]-[\textbf{w}_2-\eta\nabla F(\textbf{w}_2)]\|\leq(1+\eta\beta)\|\textbf{w}_1-\textbf{w}_2\|
\end{equation}

\noindent{\it Proof.} From the convexit of $F$,

\begin{align}
    &F(\textbf{w}_{2})\leq F(\textbf{w}_{1})+(\textbf{w}_{2}-\textbf{w}_{1})^{T}\nabla F(\textbf{w}_{1})\\
    &F(\textbf{w}_{1})\leq F(\textbf{w}_{2})+(\textbf{w}_{1}-\textbf{w}_{2})^{T}\nabla F(\textbf{w}_{2}).
\end{align}

Now summing the inequalities,

\begin{equation}
    (\textbf{w}_{2}-\textbf{w}_{1})^{T}(\nabla F(\textbf{w}_{2})-\nabla F(\textbf{w}_{1}))\geq 0.
\end{equation}

By using the $\beta$-smoothness outlined in Assumption 1,
\begin{align}
    &\|[\textbf{w}_{1}\eta\nabla F(\textbf{w}_{1}]-[\textbf{w}_{2}-\eta\nabla F(\textbf{w}_{2})]\|^{2}\\
    &=\|\textbf{w}_{1}-\textbf{w}_{2}\|^2+\eta^2\|\nabla F(\textbf{w}_{1})-\nabla F(\textbf{w}_{2})\|^2\\
    &-2[\textbf{w}_{2}-\textbf{w}_{1}][\nabla F(\textbf{w}_{2})-\eta\nabla F(\textbf{w}_{1})]\\
    &\leq(1+(\eta\beta)^2)\|\textbf{w}_{1}-\textbf{w}_{2}\|^2.
\end{align}
The result of the lemma follow accordingly.

\subsection{Proof of Lemma 7}
\noindent\textbf{Lemma 7.} \textit{Using Assumption 1, with learning rate $\eta<\frac{2}{\beta}$, for $t\in(k\tau-\Delta,(k+1)\tau-\Delta),t\neq k\tau$,}
\begin{equation}\label{Lem7LocMinCent}
    \begin{aligned}
        \E[\|\textbf{w}_i(t)-\textbf{c}_k(t)\|]&\leq\E[(1+\eta\beta)\|\textbf{w}_i(t-1)\\
        &-\textbf{c}_k(t-1)\|]\\
        &+\eta\delta_i\\
        &+\eta \Theta_i S_i\sqrt{2}\sqrt{\frac{N_i-n_i(k)}{N_in_i(k)}}
    \end{aligned}
\end{equation}

\noindent\textit{Proof.} For $t\in(k\tau-\Delta,(k+1)\tau-\Delta), t\neq k\tau$,
\begin{equation}
   \begin{aligned}
       &\textbf{w}_i(t)-\textbf{c}_k(t)\\
       &=(\textbf{w}_i(t-1)-\eta g_i(\textbf{w}_i(t-1);\xi_i(t-1))\\
       &-(\textbf{c}_k(t-1)-\eta \nabla F(\textbf{c}_k(t-1)))\\
       &=\textbf{w}_i(t-1)-\textbf{c}_k(t-1)\\
       &-\eta[\nabla F_i(\textbf{w}_i(t-1))-\nabla F_i(\textbf{c}_k(t-1))]\\
       &-\eta[\nabla F(\textbf{c}_k(t-1)-\nabla F_i(\textbf{c}_k(t-1))]\\
       &-\eta \nu_i(k)
   \end{aligned}
\end{equation}
Taking the norm and applying the triangle inequality,
\begin{equation}
    \begin{aligned}
       &\|\textbf{w}_i(t)-\textbf{c}_k\|\\
       &\leq \eta\|\nabla F_i(\textbf{w}_i(t-1))-\nabla F_i(\textbf{c}_k(t-1))\|\\
       &+\eta\|\nabla F_i(\textbf{c}_k(t-1)-\nabla F(\textbf{c}_k(t-1))\|\\
       &+\eta \|\nu_i(k)\|\\
    \end{aligned}
\end{equation}

Using Lemma 6 and Assumption 2,
\begin{equation}
    \begin{aligned}
        \|\textbf{w}_i(t)-\textbf{c}_{k}(t)\|&\leq(1+\eta\beta)\|\textbf{w}_i(t-1)-\textbf{c}_k(t-1)\|\\
        &+\eta\delta_i\\
        &+\eta\|\nu_i(k)\|
    \end{aligned}
\end{equation}

Lastly taking the expectation and applying Lemma 1,
\begin{equation}
    \begin{aligned}
        &\E[\|\textbf{w}_i(t)-\textbf{c}_k(t)\|]\\
        &\leq\E[(1+\eta\beta)\|\textbf{w}_i(t-1)-\textbf{c}_k(t-1)\|]\\
        &+\eta\delta_i\\
        &+\eta S_i\sqrt{\frac{N_i-n_i(k)}{N_in_i(k)}}
    \end{aligned}
\end{equation}

\subsection{Proof of Lemma 8}
\noindent\textbf{Lemma 8.} \textit{Under Assumption 1 with $\eta<\frac{2}{\beta}$,}
\begin{equation}\label{Lem8GlobCentAtktau}
    \begin{aligned}
       &\E[\|\textbf{w}(k\tau)-\textbf{c}_k(k\tau)\|]\\
       &\leq\alpha(k)\Delta L\eta\\
       &+(1-\alpha(k))\left[((1+\eta\beta)^{\Delta}-1)\epsilon(k)+h(\Delta,k)+\eta\Delta\sigma(k)\right]
    \end{aligned}
\end{equation}
\textit{Where $h(x,k)=\frac{\delta+\sigma(k)}{\beta}[(1+\eta\beta)^x-1]-\eta(\delta+\sigma(k))x$}\\

\noindent\textit{Proof.} By the definitions of $\textbf{w}(t)$ and $\textbf{c}_k(t)$, after some algebraic manipulations,
\begin{equation}
   \begin{aligned}
       &\textbf{w}(k\tau)=\sum_i{\rho_i\textbf{w}_i(k\tau)}\\
       &=\textbf{w}(k\tau-\Delta)\\
       &-(1-\alpha(k))\sum_{r=1}^{\Delta}\sum_i{\rho_i g_i(\textbf{w}_i(k\tau-r);\xi_{i}(k\tau-r))}\\\\
       &=\sum_i{\rho_i\textbf{w}_i(k\tau)}\\
       &-(1-\alpha(k))\eta\sum_{r=1}^{\Delta}\sum_i{\rho_i\nabla F_i(\textbf{w}_i(k\tau-r))}\\
       &-(1-\alpha(k))\eta\sum_{r=1}^{\Delta}\sum_i{\rho_i \nu_i(k)}
    \end{aligned}
\end{equation}

\noindent and

\begin{equation}
    \begin{aligned}
       \textbf{c}_k(k\tau)=\textbf{c}_k(k\tau-\Delta)-\eta\sum_{r=1}^{\Delta}\sum_i{\rho_i\nabla F_i(\textbf{c}_k(k\tau-r))}
    \end{aligned}
\end{equation}

Now we take the difference between the two previously defined terms,
\begin{equation}
    \begin{aligned}
       &\textbf{w}(k\tau)-\textbf{c}_k(k\tau)=\\
       &\eta\alpha(k)\sum_{r=1}^{\Delta}\sum_i{\rho_i \nabla F_i(\textbf{c}_k(k\tau-r))}\\
       &-(1-\alpha(k))\eta\sum_{r=1}^{\Delta}\sum_i{\rho_i[\nabla F_i(\textbf{w}_i(k\tau-r))-\nabla F_i(\textbf{c}_k(k\tau-r))]}\\
       &-(1-\alpha(k))\eta\sum_{r=1}^{\Delta}\sum_i{\rho_i \nu_i(k)},
    \end{aligned}
\end{equation}

\noindent and by taking the norm and applying the triangle inequality, we obtain

\begin{equation}
    \hspace{-3mm}\begin{aligned}
       &\|\textbf{w}(k\tau)-\textbf{c}_k(k\tau)\|\leq \eta\alpha(k)\sum_{r=1}^{\Delta}\sum_i{\rho_i\|\nabla F_i(\textbf{c}_k(k\tau-r))\|}\\
       &+(1-\alpha(k))\eta\sum_{r=1}^{\Delta}\sum_i{\rho_i\|\nabla F_i(\textbf{w}_i(k\tau-r))-\nabla F_i(\textbf{c}_k(k\tau-r))\|}\\
       &+(1-\alpha(k))\eta\sum_{r=1}^{\Delta}\sum_i{\rho_i\|\nu_i(k)\|}.
    \end{aligned}
\end{equation}

Recursively unpacking terms ending at $\textbf{w}(k\tau-\Delta)=\textbf{c}_k(k\tau-\Delta)$, taking the expectation, applying Assumption 1, and using Lemma 7,
\begin{equation}
    \begin{aligned}
       &\E[\|\textbf{w}(k\tau)-\textbf{c}_k(k\tau)\|]\leq\eta\alpha(k)\Delta L\\
       &+(1-\alpha(k))\eta\beta[\\
       &\sum_{r=1}^{\Delta}{(1+\eta\beta)^{\Delta-r}}\sum_i{\rho_i\E[\|\textbf{w}_i(k\tau-\Delta)-\textbf{w}(k\tau-\Delta)\|]}]\\
       &+(1-\alpha(k))\eta\beta\sum_{r=1}^{\Delta}\sum_{j=0}^{\Delta-r-1}{(1+\eta\beta)^j}\sum_i{\rho_i\delta_i}\\
       &+(1-\alpha(k))\eta\beta\sum_{r=1}^{\Delta}\sum_{j=0}^{\Delta-r-1}{(1+\eta\beta)^j}\sum_i{\rho_i\E[\|\nu_i(k)\|]}\\
       &+(1-\alpha(k))\eta\sum_{r=1}^{\Delta}\sum_i{\rho_i\E[\|\nu_i(k)\|]}
    \end{aligned}
\end{equation}

Lastly, we apply Lemmas 2 and 5 and use Assumption 2 to conclude that
\begin{equation}
    \begin{aligned}
       &\E[\|\textbf{w}(k\tau)-\textbf{c}_k(k\tau)\|]\\
       &\leq\eta\alpha(k)\Delta L\\
       &+(1-\alpha(k))\eta\beta\epsilon(k)\sum_{r=1}^{\Delta}{(1+\eta\beta)^{\Delta-r}}\\
       &+\frac{\delta+\sigma(k)}{\beta}(1-\alpha(k))\eta\beta\sum_{r=1}^{\Delta}{[(1+\eta\beta)^{\Delta-r}-1]}\\
       &+(1-\alpha(k))\eta\Delta\sigma(k),
    \end{aligned}
\end{equation}
with algebraic simplifications leading to the result of the lemma described above.

\subsection{Proof of Proposition 1}
\noindent\textbf{Proposition 1.} \textit{Under Assumption 1 with $\eta<\frac{2}{\beta}$,}
\begin{equation}\label{Prop1GlobCentDiff}
    \begin{aligned}
       &\E[\|\textbf{w}((k+1)\tau-\Delta)-\textbf{c}_k((k+1)\tau-\Delta)\|\\
       &\leq(1-\alpha(k))\epsilon(k)([1+\eta\beta]^{\tau}-1)\\
       &+(1-\alpha(k))h(\tau,k)+\alpha(k) h(\tau-\Delta,k)\\
       &+\alpha(k)\eta\Delta L[1+\eta\beta]^{\tau-\Delta}\\
       &+\eta\sigma(k)[\tau-\alpha(k)\Delta]\triangleq\psi(\alpha(k),k)
    \end{aligned}
\end{equation}

\noindent\textit{Proof.} Let $t\in(k\tau-\Delta,(k+1)\tau-\Delta]$. Using \eqref{SGDstep},
\begin{equation}
    \begin{aligned}
       \textbf{w}_i=\alpha_t(k)\textbf{w}(k\tau-\Delta)&+(1-\alpha_t(k))[\textbf{w}_i(t-1)\\
                                                    &-\eta g_i(\textbf{w}_i(t-1);\xi_{i}(t-1))]
    \end{aligned}
\end{equation}
\begin{equation}
    \textbf{c}_k(t)=\textbf{c}_k(t-1)-\eta\nabla F(\textbf{c}_k(t-1))
\end{equation}

Since 
\begin{equation}
    \textbf{c}_k(k\tau-1)=\textbf{w}(k\tau-\Delta)-\eta\sum_{r=0}^{\Delta-2}{\nabla F(\textbf{c}_k(k\tau-\Delta+r}))
\end{equation}
it follows that (by taking $\sum_{i}{\rho_i\textbf{w}_i}$) and expanding $g_i$ into its gradient and noise,
\begin{equation}
    \begin{aligned}
        &\textbf{w}(t)-\textbf{c}_{k}(t)\\&=(1-\alpha_t(k)[\textbf{w}(t-1)-\textbf{c}_k(t-1)]\\
        &-(1-\alpha_t(k))\eta\sum_i{\rho_i[\nabla F_i(\textbf{w}_i(t-1))-\nabla F_i(\textbf{c}_k(t-1))]}\\
        &-(1-\alpha_t(k))\eta\sum_i{\rho_i \nu_i(k)}\\
        &+\eta\alpha_t(k)\sum_{r=0}^{\Delta-1}{\nabla F(\textbf{c}_k(k\tau-\Delta+r))}
    \end{aligned}
\end{equation}

Applying the triangle inequality to the norm and applying Assumption 1 and Lemma \ref{Lem6OnePlusEtaBeta},
\begin{equation}
    \begin{aligned}
       &\|\textbf{w}(t)-\textbf{c}_k(t)\|\\&\leq(1-\alpha_t(k))\|\textbf{w}(t-1)-\textbf{c}_k(t-1)\|\\
       &(1-\alpha_t(k))\eta\beta\sum_i{\rho_i\|\textbf{w}_i(t-1)-\textbf{c}_k(t-1)\|}\\
       &+\alpha_t(k)\eta\Delta L\\
       &+(1-\alpha_t(k))\eta\sum_i{\rho_i \|\nu_i(k)\|}
    \end{aligned}
\end{equation}

For $t\in[k\tau-\Delta,k\tau-1]$, where $\alpha_t(k)=0$, and using $\textbf{c}_k(k\tau-\Delta)=\textbf{w}(k\tau-\Delta)$
\begin{equation}
    \begin{aligned}
    \|\textbf{w}(t)-\textbf{c}_k(t)\|&\leq\eta\beta\sum_{\ell=k\tau-\Delta}^{t-1}\sum_i{\rho_i\|\textbf{w}_i(\ell)-\textbf{c}_k(\ell)\|}\\
    &+\eta\sum_{\ell=k\tau-\Delta}^{t-1}\sum_i{\rho_i\|\nu_i(k)\|}
    \end{aligned}
\end{equation}

And for $t\in[k\tau,(k+t)\tau-\Delta]$, with $\alpha_{k\tau}(k)=\alpha(k)$, $\alpha_t(k)=0,\forall t>k\tau$
\begin{equation}
    \begin{aligned}
       \|\textbf{w}(t)-\textbf{c}_k(t)\|&\leq(1-\alpha(k))\eta\beta\sum_{\ell=k\tau-\Delta}^{k\tau-1}\sum_i{\rho_i}{\|\nu_i(k)\|}\\
       &+\eta\beta\sum_{\ell=k\tau}^{t-1}\sum_i{\rho_i\|\textbf{w}_i(\ell)-\textbf{c}_k(\ell)\|}\\
       &+\alpha(k)\eta\Delta L\\
       &+(1-\alpha(k))+\eta\sum_{\ell=k\tau-\Delta}^{k\tau-1}\sum_i{\rho_i\|\nu_i(k)\|}\\
       &+\eta\sum_{\ell=k\tau}^{t-1}\sum_i{\rho_i\|\nu_i(k)\|}
    \end{aligned}
\end{equation}

Which that implies that for $t=(k+1)\tau-\Delta$, by taking the expectation and applying Lemma 2 and Assumption 1,
\begin{equation}
    \begin{aligned}\label{Prop1NeedSum}
       &\E[\|\textbf{w}((k+1)\tau-\Delta)-\textbf{c}_k((k+1)\tau-\Delta)\|]\\
       &\leq(1-\alpha(k))\eta\beta\sum_{\ell=k\tau-\Delta}^{k\tau-1}\sum_i{\rho_i\E[\|\textbf{w}_i(\ell)-\textbf{c}_k(\ell)\|]}\\
       &+\eta\beta\sum_{\ell=k\tau}^{(k+1)\tau-\Delta-1}\sum_i{\rho_i\E[\|\textbf{w}_i(\ell)-\textbf{c}_k(\ell)\|]}\\
       &+\alpha(k)\eta\Delta L\\
       &+(1-\alpha(k))\eta\sum_{\ell=k\tau-\Delta}^{k\tau-1}\sigma(k)\\
       &+\eta\sum_{\ell=k\tau}^{(k+1)\tau-\Delta-1}\sigma(k)\\
    \end{aligned}
\end{equation}
With everything else solved for, the term $\sum_{i}\rho_{i}[\|\textbf{w}_i(\ell)-\textbf{c}_k(\ell)\|]$, will now be derived, beginning with
\begin{equation}
    \begin{aligned}
       &\textbf{w}_i(\ell)-\textbf{c}_k(\ell)\\
       &=(1-\alpha_\ell(k))[\textbf{w}_i(\ell-1)-\textbf{c}_k(\ell-1)]\\
       &-\eta(1-\alpha_\ell(k))[\nabla F_i(\textbf{w}_i(\ell-1))-\nabla F_i(\textbf{c}_k(\ell-1))]\\
       &-(1-\alpha_\ell(k))\eta[\nabla F_i(\textbf{c}_k(\ell-1))-\nabla F(\textbf{c}_k(\ell-1))]\\
       &+\alpha_\ell(k)\eta\sum_{r=0}^{\Delta-1}\nabla F(\textbf{c}_k(k\tau-\Delta+r))\\
       &-(1-\alpha_\ell(k))\eta \nu_i(k)
    \end{aligned}
\end{equation}
Applying $\sum_{i}\rho_i$ and taking the norm,
\begin{equation}
    \begin{aligned}
       &\sum_i{\rho_i\|\textbf{w}_i(\ell)-\textbf{c}_k(\ell)\|}\\
       &\leq(1-\alpha_\ell(k))[(1+\eta\beta)\sum_i{\rho_i\|\textbf{w}_i(\ell-1)-\textbf{c}_k(\ell-1)\|}]\\
       &+(1-\alpha_\ell(k))\eta\delta\\
       &+\alpha_\ell(k)\eta\Delta L\\
       &+(1-\alpha_\ell(k))\eta\sum_i{\rho_i\|\nu_i(k)\|}\\\\
       &=(1-\alpha_\ell(k))[(1+\eta\beta)\sum_i{\rho_i\|\textbf{w}_i(\ell-1)-\textbf{c}_k(\ell-1)\|}]\\
       &+(1-\alpha_\ell(k))\eta(\delta+\sum_i{\rho_i\|\nu_i(k)\|})\\
       &+\alpha_\ell(k)\eta\Delta L
    \end{aligned}
\end{equation}

Following a similar approach to dividing the time interval into separate parts, we first begin with the period $\ell\in[k\tau-\Delta,k\tau-1],\alpha_\ell(k)=0$,
\begin{equation}
    \begin{aligned}
       &\sum_i{\rho_i\|\textbf{w}_i(\ell)-\textbf{c}_k(\ell)\|}\\
       &\leq(1+\eta\beta)\sum_i{\rho_i\|\textbf{w}_i(\ell-1)-\textbf{c}_k(\ell-1)\|}\\
       &+\eta(\delta+\sum_i{\rho_i\|\nu_i(k)\|})\\
    \end{aligned}
\end{equation}

Recursively unpacking the first term and using the fact that $\textbf{w}(k\tau-\Delta)=\textbf{c}_k(k\tau-\Delta)$, 
\begin{equation}
    \begin{aligned}
       &\sum_i{\rho_i\|\textbf{w}_i(\ell)-\textbf{c}_k(\ell)\|}\\
       &\leq[1+\eta\beta]^{\ell-(k\tau-\Delta)}\sum_i{\rho_i\|\textbf{w}_i(k\tau-\Delta)-\textbf{w}(k\tau-\Delta)\|}\\
       &+(\delta+\sum_i{\rho_i\|\nu_i(k)\|})\frac{[1+\eta\beta]^{\ell-(k\tau-\Delta)}-1}{\beta}\\
    \end{aligned}
\end{equation}
Taking the expectation and using Lemmas 2 and 5,
\begin{equation}
    \begin{aligned}
       &\sum_i{\rho_i\E[\|\textbf{w}_i(\ell)-\textbf{c}_k(\ell)\|}\\
       &\leq\epsilon(k)[1+\eta\beta]^{\ell-k\tau+\Delta}\\
       &+(\delta+\sigma(k))\frac{[1+\eta\beta]^{\ell-k\tau+\Delta}-1}{\beta}
    \end{aligned}
\end{equation}

Similarly for $\ell\in[k\tau,(k+1)\tau-\Delta]$,
\begin{equation}
    \begin{aligned}
       &\sum_i{\rho_i\E[\|\textbf{w}_i(\ell)-\textbf{c}_k(\ell)\|]}\\
       &\leq(1-\alpha)[1+\eta\beta]^{\ell-(k\tau-\Delta)}\epsilon_k\\
       &+(1-\alpha)(\delta+\sigma(k))[1+\eta\beta]^{\ell-k\tau}\frac{[1+\eta\beta]^{\Delta}-1}{\beta}\\
       &+(\delta+\sigma(k))\frac{[1+\eta\beta]^{\ell-k\tau}-1}{\beta}\\
       &+\alpha\eta\Delta L[1+\eta\beta]^{\ell-k\tau}
    \end{aligned}
\end{equation}
Which leads to
\begin{equation}
    \begin{aligned}
       &\sum_{\ell=k\tau-\Delta}^{k\tau-1}\sum_i{\rho_i\E[\|\textbf{w}_i(\ell)-\textbf{c}_k(\ell)\|]}\\
       &\leq\epsilon(k)\frac{[1+\eta\beta]^{\Delta}-1}{\eta\beta}+\frac{h(\Delta,k)}{\eta\beta}
    \end{aligned}
\end{equation}
and
\begin{equation}
    \begin{aligned}
       &\sum_{\ell=k\tau}^{(k+1)\tau-\Delta-1}\sum_i{\rho_i\E[\|\textbf{w}_i(\ell)-\textbf{c}_k(\ell)\|]}\\
       &\leq(1-\alpha)[1+\eta\beta]^{\Delta}\epsilon(k)\frac{[1+\eta\beta]^{\tau-\Delta}-1}{\eta\beta}\\
       &+(1-\alpha)\frac{h(\tau,k)-h(\Delta,k)}{\eta\beta}+\alpha\frac{h(\tau-\Delta,k)}{\eta\beta}\\
       &+\alpha\Delta L\frac{[1+\eta\beta]^{\tau-\Delta}-1}{\beta}
    \end{aligned}
\end{equation}

\noindent The result of the lemma is thus yielded by plugging in the above into \eqref{Prop1NeedSum}:
\begin{equation}
    \begin{aligned}
       &\E[\|\textbf{w}((k+1)\tau-\Delta)-\textbf{c}_k((k+1)\tau-\Delta)\|\\
       &\leq(1-\alpha)\epsilon(k)([1+\eta\beta]^{\tau}-1)\\
       &+(1-\alpha)h(\tau,k)+\alpha h(\tau-\Delta,k)\\
       &+\alpha\eta\Delta L[1+\eta\beta]^{\tau-\Delta}\\
       &+\eta\sigma(k)[\tau-\alpha\Delta]\triangleq\psi(\alpha,k)
    \end{aligned}
\end{equation}\hspace{84mm}$\square$

\subsection{Proof of Proposition 2}
\noindent\textbf{Proposition 2.} {\it Let}
\begin{equation}
    \omega = \frac{1}{\max_{k\in\{0,...,K-1\}}\|\textbf{c}_k(k\tau-\Delta)-\textbf{w}^{\star}\|^2}.
\end{equation}

\noindent{\it Under Assumption 1, and if the following conditions are met,}
\begin{enumerate}
    \item $\eta<\frac{2}{\beta}$
    \item $T\eta\phi-\frac{L\Psi(\widehat{\bm{\alpha}})}{\varXi^2}>0$
    \item $F(\textbf{c}_k((k+1)\tau-\Delta))-F(\textbf{w}^{\star})\geq\varXi,\forall k$
    \item $F(\textbf{w}((K+1)\tau-\Delta))-F(\textbf{w}^{\star})\geq\varXi$,
\end{enumerate}

\noindent for some $\varXi>0$, we can upper-bound the convergence of {\tt StoFedDelAv} as

\begin{equation}
    F(\textbf{w}((K+1)\tau-\Delta)-F(\textbf{w}^{\star})\leq\frac{1}{T\eta\phi-\frac{L\Psi(\alpha)}{\varXi^2}},
\end{equation}

\noindent where $\Psi(\widehat{\bm{\alpha}})\triangleq\sum_{k=1}^{K}\psi(\alpha(k),k)$.
\\\\
\noindent{\it Proof.}
We consider the case $\omega<\infty$ since $\omega=\infty$ is trivially tied to $\textbf{w}((K+1)\tau-\Delta)=\textbf{c}((K+1)\tau-\Delta)=\textbf{w}^{\star}\Rightarrow F(\textbf{w}((K+1)\tau-\Delta)=F(\textbf{w}^{\star})$. Then for every $k$ and $t\in[k\tau-\Delta,(k+1)\tau-\Delta]$, we define the sub-optimality gap of the centralized GD scheme as

\begin{equation}
    \Gamma_{[k]}(t)=F(\textbf{c}_k(t))-F(\textbf{w}^{\star}),
\end{equation}
noting that $\Gamma_{[k]}(t)\geq 0,\forall k$.Since $\textbf{w}((K+1)\tau-\delta))=\textbf{c}_{[K+1]}((K+1)\tau-\Delta)$, we wish to prove that

\begin{equation}\label{eq:gamma_bnd}
    \Gamma_{[K+1]}((K+1)\tau-\Delta))^{-1}\geq T\eta\phi-\frac{L\Psi(\widehat{\bm{\alpha}})}{\varXi^2}.
\end{equation}

From the results of \cite{wang2019adaptive}'s Lemma 6, we know that

\begin{align}
    &\Gamma_{[k]}^{-1}(t+1)-\Gamma_{[k]}^{-1}(t)\geq\frac{\eta(1-(\eta\beta)/2)}{\|\textbf{c}_k(t)-\textbf{w}^{\star}\|^2}\nonumber\\
    &\geq\frac{\eta(1-(\eta\beta)/2)}{\max_{k}\|\textbf{c}_k(t)-\textbf{w}^{\star}\|^2}=\eta\omega\left(1-\frac{\eta\beta}{2}\right)=\eta\phi.
\end{align}

We therefore conclude that

\begin{align}
    &\Gamma_{[k]}^{-1}((k+1)\tau-\Delta)-\Gamma_{[k]}^{-1}(k\tau-\Delta)\label{eq:gamma_inv_k1k_diff}\\
    &=\sum_{t=k\tau-\Delta}^{(k+1)\tau-\Delta-1}\left[\Gamma_{[k]}^{-1}(t+1)-\Gamma_{[k]}^{-1}(t)\right]\geq\tau\eta\phi.
\end{align}

With this in mind, we can conclude the following:

\begin{align}
    &\sum_{k=1}^{K}\left[\Gamma_{[k]}^{-1}((k+1)\tau-\Delta)-\Gamma_{[k]}^{-1}(k\tau-\Delta)\right]\\
    &=\Gamma_{[K+1]}^{-1}((K+1)\tau-\Delta))-\Gamma_{[1]}^{-1}(\tau-\Delta)\\
    &-\sum_{k=1}^{K}\left[\Gamma_{[k+1]}^{-1}((k+1)\tau-\Delta)-\Gamma_{[k]}^{-1}((k+1)\tau-\Delta)\right]\nonumber\\
    &\geq T\eta\phi\nonumber
\end{align}

To prove \eqref{eq:gamma_bnd}, we need to show that

\begin{equation}\label{eq:gamma_sum_ineq}
    \sum_{k=1}^{K}\left[\Gamma_{[k]}^{-1}((k+1)\tau-\Delta)-\Gamma_{[k+1]}^{-1}((k+1)\tau-\Delta)\right]\leq\frac{L\Psi(\widehat{\bm{\alpha}})}{\varXi^2}.
\end{equation}

Since $\Psi(\widehat{\bm{\alpha}})=\sum_{k=1}^{K}\psi(\alpha(k),k)$, \eqref{eq:gamma_sum_ineq} is implied by

\begin{align}
    &\Gamma_{[k+1]}((k+1)\tau-\Delta)-\Gamma_{[k]}((k+1)\tau-\Delta)\label{eq:gamma_prv_bnd}\\
    &\leq\frac{L\psi(\alpha(k),k)}{\varXi^2}\Gamma_{[k]}((k+1)\tau-\Delta)\Gamma_{[k+1]}((k+1)\tau-\Delta).
\end{align}

Conditions (3) and (4) from the proposition statement imply that

\begin{align}
    &\Gamma_{[k]}((k+1)\tau-\Delta)\geq\varXi,\forall k,\\
    &\Gamma_{[K+1]}((K+1)\tau-\Delta)\geq\varXi.
\end{align}

Using \eqref{eq:gamma_inv_k1k_diff}, with $k<K-1$,

\begin{align}
    &\Gamma_{[k+1]}((k+1)\tau-\Delta)\geq\frac{\Gamma_{[k+1]}((k+2)\tau-\Delta)}{1-\tau\eta\phi\Gamma_{[k+1]}((k+2)\tau-\Delta)}\\
    &\geq\Gamma_{[k+1]}((k+2)\tau-\Delta)\geq\varXi.
\end{align}

The above statements show that \eqref{eq:gamma_prv_bnd} can be proven by showing that

\begin{equation}
    \Gamma_{[k+1]}((k+1)\tau-\Delta)-\Gamma_{[k]}((k+1)\tau-\Delta)\leq L\psi(\alpha(k),k).
\end{equation}

This is in fact the case. By combining with Proposition 1, we obtain:

\begin{align}
    &\Gamma_{[k+1]}((k+1)\tau-\Delta)-\Gamma_{[k]}((k+1)\tau-\Delta)\\
    &=F(\textbf{w}((k+1)\tau-\Delta))-F(\textbf{c}_k((k+1)\tau-\Delta))\\
    &\leq L\E[\|\textbf{w}((k+1)\tau-\delta)-\textbf{c}_k((k+1)\tau-\Delta)]\|.
\end{align}

\noindent The result of Proposition 2 directly follows.\hspace{23mm} $\square$

\subsection{Proof of Theorem 1}
\noindent\textbf{Theorem 1.}\textit{With $\eta<\frac{2}{\beta}$ and under Assumption 1.}
\begin{equation}
\begin{aligned}
    &F(\textbf{w}^{K})-F(\textbf{w}^{\star})\\
    &\leq \frac{1}{2\eta\phi T}+\sqrt{\frac{1}{(2\eta\phi T)^2}+\frac{L\Psi(\alpha)}{\eta\phi T}}+L\Psi(\widehat{\bm{\alpha}})
\end{aligned}
\end{equation}
\textit{where $\Psi(\widehat{\bm{\alpha}})=\sum_{k=1}^{K}\psi(\alpha,k)$.}\\\\
\noindent\textit{Proof.} To prove Theorem 1, we first begin by defining an auxiliary variable $\varXi^{*}>0$ given $\eta\leq\frac{1}{\beta}$ such that $T\eta\phi-\frac{L*\Psi(\widehat{\bm{\alpha}})}{\varXi^{*2}}>0$ and
$\varXi^{*}=\frac{1}{T\eta\phi-\frac{L*\Psi{\widehat{\bm{\alpha}}}}{\varXi^{*2}}}$. Solving these equations for $\varXi^{*}$ yields
\begin{equation}
    \varXi^{*}=\frac{1}{2\eta\phi T}+\sqrt{\left(\frac{1}{2\eta\phi T}\right)^2+\frac{L\Psi({\widehat{\bm{\alpha}}})}{\eta\phi T}}
\end{equation}
Letting $\varXi>\varXi^{*}$, and assuming that the conditions of Proposition 2 are satisfied, it follows that
\begin{equation}
    \begin{aligned}
    F(\textbf{w}((K+1)\tau-\Delta))-F(\textbf{w}^{\star})&<\frac{1}{T\eta\phi-\frac{L\Psi(\widehat{\bm{\alpha}})}{\varXi^{2}}}\\
    &\leq\frac{1}{T\eta\phi-\frac{L\Psi(\widehat{\bm{\alpha}})}{\varXi^{*2}}}\\
    &\Rightarrow\varXi^{*}<\varXi
    \end{aligned}
\end{equation}
This presents a contradiction with the fourth condition of Prop. 2, meaning that at least one of those conditions cannot be satisfied given $\varXi>\varXi^{*}$. The first two conditions are readily satisfied and
\begin{equation}
    \varXi>\varXi^{*}=\frac{1}{T\eta\phi-\frac{L\Psi(\widehat{\bm{\alpha}})}{\varXi^{*2}}}
\end{equation}
With either the third or fourth conditions not met, we therefore conclude that
\begin{equation}\label{eq:prop2cdt34viol}
    \begin{aligned}
        &\min{\left\{F(\textbf{w}((K+1)\tau-\Delta)),\min{F(\textbf{c}_k((k+1)\tau-\Delta))}\right\}}\\
        &-F(\textbf{w}^{\star})\leq\varXi^{\star}
    \end{aligned}
\end{equation}
Therefore, using Prop. 1 and noting that $\psi(\alpha(k),k)$ is increasing as a function of $k$,
\begin{align}
    &F(\textbf{w}((k+1)\tau-\Delta))\leq F(\textbf{c}_k((k+1)\tau-\Delta))\\\nonumber
    &+\abs{F(\textbf{w}((k+1)\tau-\Delta))- F(\textbf{c}_k((k+1)\tau-\Delta))}\\\nonumber
\end{align}
Taking the norm and expectation,
\begin{align}
    &\leq F(\textbf{c}_k((k+1)\tau-\Delta))\\\nonumber
    &+L\E\left[\|\textbf{w}((k+1)\tau-\Delta)-\textbf{c}_k((k+1)\tau-\Delta)\|\right]\\
    &\leq F(\textbf{c}_k((k+1)\tau-\Delta))+L\psi(\alpha(k),k)\\
    &\leq F(\textbf{c}_k((k+1)\tau-\Delta)) +\Psi(\widehat{\bm{\alpha}})
\end{align}
Implying
\begin{align}
    &\min_{k}\{F(\textbf{c}_k((k+1)\tau-\Delta))\}\\\nonumber
    &\geq\min_{k}{F(\textbf{w}((k+1)\tau-\Delta))-L\Psi(\widehat{\bm{\alpha}})}
\end{align}
Using the result of \eqref{eq:prop2cdt34viol},
\begin{equation}
    \min_{k\leq K}{F(\textbf{w}((k+1)\tau-\Delta))}-L\Psi(\widehat{\bm{\alpha}})-F(\textbf{w}^{*})\leq\varXi^{*}
\end{equation}
with the theorem following as a direct consequence.\hspace{9mm}$\square$


%

\end{document}